\documentclass[journal]{IEEEtran}
\usepackage{cite,amsmath,graphicx,amssymb,cite,amsthm,multirow}
\usepackage[ruled,lined,boxed,commentsnumbered]{algorithm2e}
\usepackage{algorithmic}
\usepackage[table]{xcolor}

\newtheorem{theorem}{Theorem}[section]

\newtheorem{lemma}[theorem]{Lemma}

\newcommand{\specialcell}[2][c]{%
  \begin{tabular}[#1]{@{}c@{}}\vspace{-.15in}#2\end{tabular}}
  
\newcounter{ALC@tempcntr}

\graphicspath{{Figures/}}

\begin{document}

\title{Particle Filtering with Invertible Particle Flow}

\author{Yunpeng~Li,~\IEEEmembership{Student Member,~IEEE,}       and~Mark~Coates,~\IEEEmembership{Senior~Member,~IEEE}
\thanks{Y. Li and M. Coates are with the Department of Electrical and Computer
Engineering, McGill University, Montr\'eal, Qu\'ebec, Canada,
e-mail: yunpeng.li@mail.mcgill.ca; mark.coates@mcgill.ca.}
}


\maketitle

%
\begin{abstract}

A key challenge when designing particle filters in high-dimensional
  state spaces is the construction of a proposal distribution that is
  close to the posterior distribution.  Recent advances in particle
  flow filters provide a promising avenue to avoid weight degeneracy;
  particles drawn from the prior distribution are migrated in the
  state-space to the posterior distribution by solving partial
  differential equations.  Numerous particle flow filters have been
  proposed based on different assumptions concerning the flow
  dynamics. Approximations are needed in the implementation of all of
  these filters; as a result the particles do not exactly match a
  sample drawn from the desired posterior distribution. Past efforts
  to correct the discrepancies involve expensive calculations of
  importance weights. In this paper, we present new filters which
  incorporate deterministic particle flows into an encompassing
  particle filter framework. The valuable theoretical guarantees
  concerning particle filter performance still apply, but we can
  exploit the attractive performance of the particle flow methods.
  The filters we describe involve a computationally efficient weight
  update step, arising because the embedded particle flows we design
  possess an invertible mapping property.  We evaluate the proposed
  particle flow particle filters' performance through numerical
  simulations of a challenging
  multi-target multi-sensor tracking scenario and complex
  high-dimensional filtering examples.
\end{abstract}


%
\IEEEpeerreviewmaketitle

\section{Introduction}
\label{sec:intro}

Particle filters are a family of Monte Carlo algorithms developed to
solve the filtering problem of sequentially estimating a state
variable. Particles (state samples) and their associated weights are
advanced through time to approximate the filtering distributions of
interest. The bootstrap particle filter (BPF) draws particles from the
prior distribution and updates the weight of each particle using the
likelihood of the latest measurement~\cite{gordon1993}. When the state
dimension is high or when measurements are highly informative, the 
majority of particles drawn from the prior distribution will be in regions 
with very low likelihood, leading to 
negligible weights for most particles in the BPF. As the
Monte Carlo approximation of the posterior distribution
is dominated by a few particles, this weight degeneracy issue results in 
a poor representation of the posterior distribution~\cite{bickel2008,bunch2016}.

The ``optimal'' proposal distribution minimizes the variance of the importance  weights\cite{doucet2000b} but is rarely possible to sample from. More advanced particle filters construct efficient proposal distributions by approximating the optimal proposal distribution
\cite{doucet2000b,cornebise2008}. 
The auxiliary particle filter (APF)~\cite{pitt1999} introduces an auxiliary variable to more effectively sample particles, taking information from the new measurement into account. The Rao-Blackwellised particle filters~\cite{doucet2000a}
reduce the variance of Monte Carlo estimates by marginalizing some states analytically. The unscented particle filter~\cite{vanDerMerwe2000} approximates
the optimal proposal distribution using the unscented transformation.
Although these approaches can be effective in some settings, particle filtering in high-dimensional spaces remains a challenging task and most conventional particle filters perform
poorly~\cite{bengtsson2008,snyder2008,beskos2014a}.
Several directions have been explored in order to address the
challenge in high-dimensional filtering and in settings where the measurements are highly informative. We provide a more detailed discussion of the research
contributions most related to our work in
Section~\ref{sec:related_work}. One method that is promising and exhibits good performance in  very high dimensions is the equivalent weights particle filter~\cite{vanLeeuwen2010,ades2015}. In its basic form, it sacrifices the statistical consistency of the conventional particle filter but ensures that a large number of particles always have substantial weight, thus avoiding degeneracy.
Another approach involves separating the
state space through factorization or
partitioning~\cite{djuric2007,djuric2013,rebeschini2015,beskos2017}. These techniques are promising, but rely on identifying a suitable
factorization of the conditional posterior, so their applicability is
restricted. A more general approach involves the incorporation of
Markov Chain Monte Carlo (MCMC) methods within the particle
filters~\cite{berzuini1997,gilks2001,godsill2001,musso2001,golightly2006,beskos2014a,maroulas2012,kang2014,brockwell2010,septier2016}.

Although they can be effective in allowing particle filters to operate
in high-dimensional state spaces, MCMC methods are almost always
computationally expensive and their inclusion can render real time
filtering impossible. An alternative set of methods, labelled
``progressive Bayesian update'', ``homotopy'' or ``particle flow'',
can offer similar performance without the same computational
requirements, but this comes at the cost of a more limited theoretical
understanding. A framework for performing a progressive Bayesian
update was introduced in~\cite{hanbeck2003}. In the context of
particle filters, particles are ``migrated'' to represent the
posterior distribution; the importance sampling step is eliminated. In
a series of papers
\cite{daum2007,daum2008,daum2009,daum2010a,daum2010b,daum2011,daum2012a,daum2013a,daum2014a,daum2014b},
Daum et al.\ link the log prior (or predictive posterior) and the log posterior distribution via a homotopy and derive partial differential equations (PDE) in order to guide particles to flow from the prior (or predictive posterior) towards the posterior distribution.  Recently, Khan et al.~\cite{khan2014} and de Melo et al.~\cite{deMelo2015} have proposed alternative approaches. 
Although numerous such {\em particle flow filters} have been
developed, most solutions are analytically intractable. One important
exception is the ``exact'' particle flow~\cite{daum2010a} filter,
developed under the assumption that the measurement model is linear
and both the prior and posterior distributions are Gaussian.

For all particle flow filters, discretization is needed during
implementation to numerically solve the derived PDEs. Particles
advance through a large number of small steps. The truncation errors
accumulated through the numerical integration steps, as well as
approximations (Gaussianity, local linearity) used in various other stages of implementation,
can lead to particles deviating from the true posterior distribution.

An alternative perspective is to view the particle flow filters as a
way to generate a proposal distribution close to the posterior for use
by an encompassing particle filter. With this approach, we exploit the
ability of particle flow filters to move particles into regions where
the posterior is significant, but the extensive
theoretical understanding of sampling-based particle filters still
applies---convergence rates and large
deviations bounds are available (see~\cite{crisan2002} for examples). Several filters adopting this strategy have been recently
proposed~\cite{reich2013,bunch2016,deMelo2015,heng2015,li2015}.
Although elegant, most approaches are computationally very expensive.

In this paper, we present particle filtering algorithms that employ a
modified deterministic particle flow approach to construct a proposal
distribution that closely matches the target posterior.
Our main contributions compared to the
state-of-the-art work in this domain include:
(i) we modify the particle flow procedures
so that they constitute invertible mappings, which
allow much more efficient importance weight evaluation
than state-of-the-art particle filters that use particle flows;
(ii) we provide the proof of the invertible mapping
property of the designed particle flows;
(iii) we demonstrate the performance of particle filters with invertible  particle flows in very
challenging filtering scenarios with either highly informative measurements or high-dimensional state space.

Preliminary results
were published in abbreviated forms in conference
papers~\cite{li2016a,li2016b}. This paper provides a more detailed description
of the proposed algorithms and related methods, a more complete
performance evaluation with added high-dimensional filtering
simulations, as well as the proof of the invertible mapping property. 

The remainder of the paper is organized as follows. We provide a more
detailed discussion of related work in Section~\ref{sec:related_work}.
We present the problem statement in Section~\ref{sec:ps},
followed by a brief review of particle 
flow methods in Section~\ref{sec:flow}.
We describe the proposed particle flow particle filters
and prove the invertible mapping property in
Section~\ref{sec:algorithm}. 
Section~\ref{sec:simulation} details the 
simulation setup and presents results.
The paper's contributions and results are summarized and discussed in Section~\ref{sec:conclusion}.

\subsection{Related Work}
\label{sec:related_work}

Over the past two decades, several directions have been explored for
improving particle filtering performance in high-dimensional settings
or when measurements are highly informative. One popular class of methods
involve using the extended Kalman filter (EKF), the unscented Kalman filter
(UKF)~\cite{julier1997} or the ensemble Kalman filter (EnKF)~\cite{evensen2003} to generate the proposal distribution~\cite{vanDerMerwe2000,vanLeeuwen2015}.
These methods can be very effective for certain nonlinear models, but performance can deteriorate with non-Gaussian models. Other methods involve factorization or partitioning of the state space. In the {\em multiple
 particle filtering} approach~\cite{djuric2007,djuric2013}, particle
filters are executed in parallel on low-dimensional subspaces that
partition the full state space. The filters share information
with one another to approximate filtering in the full state space.
Rebeschini et al.\ proposed a similar strategy called the block
particle filter~\cite{rebeschini2015}, acknowledging that the blocking
(partitioning) process introduces a bias that is difficult to
quantify. The {\em space-time particle filters}~\cite{beskos2017}
also rely on factorization of the conditional posterior, but Beskos et
al.\ demonstrate that the filters provide consistent estimates as the
number of particles increases. Although they are promising, the need
to identify an effective factorization means that these algorithms are
not generally applicable.

A different direction involves the incorporation of Markov Chain Monte
Carlo (MCMC) methods within the particle filters. The first well-known
technique in this category is the resample-move
algorithm~\cite{gilks2001}, which applies MCMC after the resampling
step in order to diversify the particles. More recent techniques that
adopt a similar approach include~\cite{maroulas2012,kang2014}. These
latter approaches modify the stochastic differential equation (SDE)
which describes the state dynamics, in order to facilitate the MCMC
moves and render them more efficient.

One problem associated with performing MCMC after resampling
is that often very few particles are replicated after resampling in
the high-dimensional space. Many MCMC iterations may then be needed to
produce a set of particles that represent a reasonably independent
sample from the target posterior.  For this reason, Sequential Markov
chain Monte Carlo (SMCMC) methods avoid the resampling step and use
MCMC to generate samples directly from a proposal
distribution~\cite{berzuini1997,golightly2006,brockwell2010,septier2016}.
The identification of effective MCMC kernels often
then becomes the major challenge. In~\cite{septier2016}, MCMC kernels based
on Langevin diffusion or Hamiltonian dynamics have been proposed
to more efficiently traverse a high-dimensional space.  One of the
most effective algorithms, referred to as the SmHMC algorithm, is based on the Manifold Hamiltonian Monte Carlo kernel~\cite{septier2016}.  One
limitation of the SmHMC method is that it requires the target distribution to be log-concave. Solutions to this limitation
are proposed in~\cite{septier2016}, but they require either analytically tractable expected values of negative Hessians or careful parameter tuning. In comparison, the particle flow particle filter we describe has a milder assumption on the smoothness of the measurement function.

Another way to mitigate weight degeneracy is to approach the true
posterior density from the tractable prior density via intermediate
densities, in a similar vein to simulated annealing. Such approaches
were proposed in~\cite{godsill2001,musso2001}. Beskos et al.~\cite{beskos2014a} illustrated
how bridging densities and resampling could be used in conjunction
with the sequential Monte Carlo sampler to improve the performance of
particle filters in high dimensions. Theoretical results
in~\cite{beskos2014a} suggested that the approach successfully avoided
the degeneracy and eliminated the need for exponential growth in
the number of particles with respect to the state dimension. 

Our work is most closely related to other techniques that combine
particle flow or transport and particle
filtering~\cite{reich2013,heng2015,bunch2016,deMelo2015}. The Gaussian particle
flow importance sampling (GPFIS) algorithm developed in
\cite{bunch2016} uses approximate Gaussian flows to sample from
non-Gaussian models. There is a non-trivial weight correction after every
iteration of the particle flow and as a result the algorithm is
computationally demanding. The particle flow
particle filter (PF-PF) we describe performs an exact weight update
instead of the approximate weight updates in the
GPFIS~\cite{bunch2016}.
The stochastic particle flow technique in
\cite{deMelo2015} builds upon stationary solutions to the
Fokker-Planck equation to compose Gaussian mixtures to approximate the
posterior. The performance reported in \cite{deMelo2015} is
impressive, but the computational requirements are significant. The optimal transport method-based approaches proposed in
\cite{reich2013,heng2015} are similar in spirit to the particle flow approaches in that they involve particle transport (i.e., flow or migration)
rather than importance sampling. The filter proposed in~\cite{reich2013} incorporates weight correction after application
of a transport map, so it inherits the theoretical properties of particle filters.
The Gibbs flow approach in~\cite{heng2015}
requires numerical integrations of probability densities
in state updates of each particle at each intermediate flow step,
performed for each state dimension. So, the computational cost is very high,
especially for high dimensional filtering scenarios.

The approach we propose may be interpreted as a guided sequential
Monte Carlo (GSMC) method as outlined in~\cite{reich2013}.
The GSMC methods use deterministic transport maps
to generate the proposal density. The importance weight evaluation
step in the GSMC methods requires the calculation of the determinant
of the Jacobian matrix of the transport map. This can be a major
impediment for complex transport maps. As explained in~\cite{reich2013}, if we impose an extra assumption on the coupling that the  transport map couples the prior and the posterior {\em exactly}, then the calculation of the determinant of the transport map can be  avoided. Maps that achieve the exact coupling are identified in~\cite{reich2013} for the special cases when the prior and posterior are Gaussians or mixtures of Gaussians; the maps can be implemented using the ensemble Kalman filter.
In this paper, we effectively identify an alternative deterministic transport map using a modified exact Daum-Huang particle flow~\cite{daum2010a}
and present complete routines to construct the proposal
density using the modified invertible particle flow.
The simple structure of the map allows efficient computation of the 
weight update. In one version of the proposed particle
flow particle filter, we avoid evaluation of the determinant because
the same map is applied to all particles; in the other we can evaluate
it analytically for a relatively small computational overhead because
of the structure of the map (the repeated application of affine
transformations).

\section{Problem Statement}
\label{sec:ps}

The nonlinear filtering task we address involves tracking the marginal posterior
distribution $p(x_k|z_{1:k})$, where $x_k$ is the state of a system at time $k$ and $z_{1:k} = \{z_1,\ldots,z_k\}$ is a sequence of measurements collected up to time step $k$.
The state evolution and measurements are described by the following model:
\begin{align}
x_0 &\sim p_0(x)\,\,,\\
x_k &= g_k(x_{k-1},v_k)\quad \mbox{for~} k\geq 1 \label{eq:dynamic}\,,\\
z_k &= h_k(x_k,w_k)\quad \mbox{for~} k\geq 1 \label{eq:measurement}\,.
\end{align}
Here $p_0(x)$ is an initial probability density function,
$g_k: \mathbb{R}^{d} \times \mathbb{R}^{d'} \rightarrow
\mathbb{R}^{d}$ is the state-transition function of the unobserved state
$x_k \in \mathbb{R}^{d}$,
$z_k \in \mathbb{R}^{S}$ is the measurement generated from the state
$x_k$ through a potentially nonlinear measurement model
$h_k:\mathbb{R}^{d} \times \mathbb{R}^{S'} \rightarrow \mathbb{R}^{S}$.
$v_k \in \mathbb{R}^{d'}$ is the process noise and
$w_k\in \mathbb{R}^{S'}$ is the measurement noise.
We assume that $g_k(\cdot,0)$ is bounded, and $h_k(\cdot,0)$ is a $C^1$
function~\cite{warner1983}, i.e.
$h_k(\cdot,0)$ is differentiable everywhere and its
derivatives are continuous.

\section{Particle flows}
\label{sec:flow}

In this section we briefly review how deterministic particle flow can be used to address the nonlinear filtering problem. Suppose that we have a set of $N_p$ particles
$\{x_{k-1}^i\}_{i=1}^{N_p}$
approximating the posterior distribution
at time $k{-}1$. After propagating particles using
the dynamic model, we obtain particles
$\{\tilde{x}_k^i\}_{i=1}^{N_p}$ that represent
the predictive posterior distribution at time $k$. Particle flow is then used to migrate the particles so that they approximate the posterior distribution at time $k$.

We can model the particle flow as
a background stochastic process
$\eta_{\lambda}$ in a pseudo time interval
$\lambda \in [0, 1]$.
To simplify notation, we do not include the time index $k$ in the
following description of the stochastic process, as the particle flow only
concerns particle migration between two adjacent time steps.
We denote by $\eta_{\lambda}^i$
the stochastic process's $i$-th realization,
and set $\eta_{0}^i = \tilde{x}_k^i$, for $i = 1, 2, \ldots, N_p$.

The {\em zero diffusion} particle flow filters\cite{daum2007,daum2008,daum2009,daum2010a,daum2010b,daum2011} involve no
random displacements of particles; the flows are deterministic.
The trajectory of $\eta_{\lambda}^i$ for realization $i$
follows the ordinary differential equation (ODE):
\begin{align}
\dfrac{d\eta_\lambda^{i}}{d\lambda} = \zeta(\eta_\lambda^{i},\lambda)\,,
\label{eq:ODE_particle_flow}
\end{align} 
where $\zeta: \mathbb{R}^d \rightarrow \mathbb{R}^d$
is governed by the Fokker-Planck equation and additional
flow constraints~\cite{daum2010b}.
The Fokker-Planck equation with zero diffusion is given by
\begin{align}
&\hspace{-0.75cm}\dfrac{\partial p(\eta_\lambda^{i},\lambda)}{\partial \lambda} = -\mbox{div}(p(\eta_\lambda^{i},\lambda)\zeta(\eta_\lambda^{i},\lambda))\nonumber\\
&=-p(\eta_\lambda^{i},\lambda)\mbox{div}(\zeta(\eta_\lambda^{i},\lambda))-\dfrac{\partial p(\eta_\lambda^{i},\lambda)}{\partial \eta_\lambda^{i}}\zeta(\eta_\lambda^{i},\lambda)\,,
\label{eq:fokker_planck_zero_diffusion}
\end{align}
where $p(\eta_{\lambda}^i,\lambda)$ is the probability
density of $\eta_{\lambda}^i$ at time $\lambda$ of the flow.

By imposing different constraints on the flow, Equation~\eqref{eq:fokker_planck_zero_diffusion}
can lead to a variety of particle flow filters.
However, very few are analytically tractable.
One exception is when the predictive posterior
and the likelihood distributions are both Gaussian and the measurement
model is linear, i.e.,
$\eta_0^i \sim N(\bar{\eta}_0, P),
z = h(\eta_\lambda^i, w) \sim N(H\eta_\lambda^i, R)$.
The predictive covariance $P$
and the measurement covariance $R$ are both
positive definite.
$H$ is called the measurement matrix.
Since we drop the time index $k$ in this section,
we use $z$ to denote the measurement available at time step $k$.

\subsection{The exact Daum and Huang filter}

The flow trajectory in the resultant exact Daum and
Huang (EDH) filter~\cite{daum2010a} becomes:
\begin{align}
\zeta(\eta_\lambda^{i},\lambda)
= A(\lambda)\eta_\lambda^{i}+b(\lambda)\,,
\label{eq:EDH_mapping}
\end{align} 
where 
\begin{align}
A(\lambda)&=-\dfrac{1}{2}PH^T(\lambda HPH^T+R)^{-1}H,\label{eq:EDH_A_linear}\quad \\
b(\lambda)&=(I+2\lambda A(\lambda))[(I+\lambda A(\lambda))PH^TR^{-1}z
+A(\lambda)\bar{\eta}_0].
\label{eq:EDH_b_linear}
\end{align}

For nonlinear observation models, a linearization of the model
is performed at the mean of the intermediate distribution,
$\bar{\eta}_{\lambda}$, to construct $H(\lambda)$:
\begin{align}
H(\lambda)=\dfrac{\partial h(\eta,0)}{\partial \eta}\bigg|_{\eta=\bar{\eta}_{\lambda}} \,.
\end{align}
This is the $S \times d$ Jacobian matrix evaluated at $\bar{\eta}_{\lambda}$. A slight change must be made to the expressions of the flow, as presented in~\cite{ding2012}:
\begin{align}
A(\lambda)=&-\dfrac{1}{2}PH(\lambda)^T(\lambda H(\lambda)PH(\lambda)^T+R)^{-1}H(\lambda),\label{eq:EDH_A_nonlinear}\quad\\
b(\lambda)=&(I+2\lambda A(\lambda))[(I+\lambda A(\lambda))PH(\lambda)^TR^{-1}(z-e(\lambda))\nonumber \\
&+A(\lambda)\bar{\eta}_0]\,,
\label{eq:EDH_b_nonlinear}
\end{align}
where $e(\lambda) = h(\bar{\eta}_{\lambda},0) - H(\lambda)\bar{\eta}_{\lambda}$.

\subsection{The localized exact Daum and Huang filter}

The localized exact Daum and Huang filter (LEDH)~\cite{ding2012}
linearizes the system and updates the drift term for each individual particle. For the $i$-th particle, the drift term
\begin{equation}
\zeta(\eta_{\lambda}^i,\lambda) = A^i(\lambda)\eta_\lambda^{i}+b^i(\lambda) \,,
\label{eq:LEDH_mapping}
\end{equation}
where
\begin{align}
A^i(\lambda)=&-\dfrac{1}{2}PH^i(\lambda)^T(\lambda H^i(\lambda)PH^i(\lambda)^T+R)^{-1}H^i(\lambda),\label{eq:LEDH_A_nonlinear}\quad \\
b^i(\lambda)=&(I+2\lambda A^i(\lambda))[(I+\lambda A^i(\lambda))PH^i(\lambda)^TR^{-1}(z-e^i(\lambda))\nonumber\\
&+A^i(\lambda)\bar{\eta}_0].
\label{eq:LEDH_b_nonlinear}
\end{align}
Here $H^i(\lambda)=\dfrac{\partial h(\eta,0)}{\partial \eta}
\bigg|_{\eta=\eta^i_{\lambda}}$ and
$e^i(\lambda) = h(\eta^i_{\lambda},0) - H^i(\lambda)\eta^i_{\lambda}$.

\subsection{Numerical Implementation}

In the implementation of the exact particle flow algorithms,
discretized pseudo-time integration
is used to approximate the solution to the ODE.
Suppose that a sequence of discrete steps
are taken at $N_\lambda$ positions $[\lambda_1,
\lambda_2,\ldots,\lambda_{N_\lambda}]$,
where $0 = \lambda_0 < \lambda_1
<\ldots<\lambda_{N_\lambda}=1$.
The step size $\epsilon_j = \lambda_j-\lambda_{j-1}$
for $j=1,\ldots,N_\lambda$ can be
possibly varying, and we require
that $\sum_{j=1}^{N_\lambda}
\epsilon_j = \lambda_{N_\lambda}
-\lambda_0 = 1$.

The integral between $\lambda_{j-1}$ and $\lambda_{j}$ for $1 \leq j \leq N_\lambda$
is approximated and the functional mapping
for the EDH flow becomes 
\begin{align}
\eta_{\lambda_{j}}^{i} &= f_{\lambda_j}(\eta_{\lambda_{j-1}}^i)\nonumber\\
& = \eta_{\lambda_{j-1}}^{i}
+ \epsilon_j(A(\lambda_{j})\eta_{\lambda_{j-1}}^i
+ b(\lambda_{j}))\,.
\label{eq:discrete_update}
\end{align}
The linearization of $H(\lambda_j)$ subsequently used to
update $A(\lambda_j)$ is performed
at $\bar{\eta}_{\lambda_{j-1}}$,
which is the average of all particles $\{\eta_{\lambda_{j-1}}^i\}$
at time step $\lambda_{j-1}$.

For the LEDH, the functional mapping is
\begin{align}
\eta_{\lambda_{j}}^{i} &= f_{\lambda_j}^i(\eta_{\lambda_{j-1}}^i)\nonumber\\
& = \eta_{\lambda_{j-1}}^{i}
+ \epsilon_j(A^i(\lambda_{j})\eta_{\lambda_{j-1}}^i
+ b^i(\lambda_{j}))\,.
\label{eq:discrete_update_LEDH}
\end{align}
In the LEDH, the linearization of $H^i(\lambda_j)$ needed to
update $A^i(\lambda_j)$ is performed
at $\eta^i_{\lambda_{j-1}}$.

\section{Particle Flow with Invertible Mapping}
\label{sec:algorithm}

In the particle flow particle filtering framework we propose in this paper, the migrated particle
$\eta_1^i$ after the particle flow process
is viewed as being drawn from a proposal distribution
$q(\eta_1^i|x_{k-1}^i,z_{k})$.
In general, we cannot evaluate this proposal
distribution,
due to approximations in the filter 
implementation and the mismatch between the
model assumptions of the embedded particle
flow filter and the real scenario.

However, if the flow process defines an
invertible deterministic mapping $\eta_1^i = T(\eta_0^i;z_k,x_{k-1}^i)$ between the $i$-th particle
value before and after the flow,
we can evaluate the proposal density as follows:
\begin{align}
q(\eta_1^i|x_{k-1}^i,z_{k}) &=  \dfrac{p(\eta_0^i|x_{k-1}^i,z_{k})}{|\dot{T}(\eta_0^i;z_k,x_{k-1}^i)|}\nonumber\\
&= \dfrac{p(\eta_0^i|x_{k-1}^i)}{|\dot{T}(\eta_0^i;z_k,x_{k-1}^i)|}\,,
\label{eq:proposal}
\end{align}
where $\dot{T}(\cdot) \in
\mathbb{R}^{d\times d}$
is the Jacobian determinant of the
mapping function $T(\cdot)$ for the
$i$-th particle and $|\cdot|$ denotes the absolute value.
The mapping can be different
for each particle and can depend only on the measurements $z_{1:k}$ and the $i$-th particle's historical state values $x_{0:k-1}^i$, i.e. it cannot depend on the state values of other particles. We choose to restrict the dependence to the current measurement $z_k$ and the previous state value $x^i_{k-1}$. The first equality of~\eqref{eq:proposal}\ is
due to the invertible mapping between $\eta_0^i$ and $\eta_1^i$.  The
second holds because $\eta_0^i$ is generated solely through the
dynamic model.

We can then evaluate the importance weight of each particle
at time step $k$ as:
\begin{align}
w_{k}^{i} \propto \dfrac{p(\eta_1^{i}|x_{k-1}^{i})p(z_k|\eta_1^{i})|\dot{T}(\eta_0^i;z_k,x_{k-1}^i)|}{p(\eta_0^i|x_{k-1}^i)}w_{k-1}^i \,\,.
\label{eq:weight_update}
\end{align}
As noted in~\cite{bunch2016}, the state update during the particle
flow process is not in general an invertible mapping,
so~\eqref{eq:weight_update} does not hold. This motivated the
development of complicated weight update procedures
in~\cite{bunch2016} to approximate the importance weights.  In this
section, we propose modified particle flow procedures that possess the
invertible mapping property, which allows us to perform efficient
weight updates using~\eqref{eq:weight_update}.

\subsection{Particle Flow Particle Filtering with the LEDH flow}
\label{sec:PFPF_LEDH}

The particle flow particle filter algorithm (PF-PF) based on the LEDH
flow is
presented in Algorithm~\ref{alg:PF-PF_LEDH}. We show below that, under
certain conditions, the function constructed by the discretized
particle flow in lines 11-21 of the algorithm leads to an invertible
mapping. For the PF-PF (LEDH), we have:
\begin{align}
|\dot{T}(\eta_0^i;z_k,x_{k-1}^i)|
&=|\det(\dfrac{d\eta_1^i}{d\eta_0^i})|\nonumber\\
&=|\det(\dfrac{d[\prod_{j=1}^{N_\lambda}(I+\epsilon_jA_j^i(\lambda))]\eta_0^i}{d\eta_0^i})|\nonumber\\
&=\prod_{j=1}^{N_\lambda}|\det(I+\epsilon_jA_j^i(\lambda))|\,.
\label{eq:jac_det_LEDH}
\end{align}
where $\det(\cdot)$ denotes the determinant.
We prove at the end of Section~\ref{sec:PFPF_LEDH} that
$0<|\dot{T}(\eta_0^i;z_k,x_{k-1}^i)|<\infty$. Thus, the weight update expression for
PF-PF (LEDH) is
\begin{align}
w_{k}^{i} \propto \dfrac{p(\eta_1^{i}|x_{k-1}^{i})p(z_k|\eta_1^{i})\prod_{j=1}^{N_\lambda}|\det(I+\epsilon_jA_j^i(\lambda))|}{p(\eta_0^i|x_{k-1}^i)}w_{k-1}^i
\label{eq:weight_update_PFPF_LEDH}
\end{align}

The particle flow procedure for each particle requires
a predicted covariance estimate, i.e.,
the covariance matrix of the predictive posterior.
The predicted covariance $P$ can be obtained by using the
Kalman covariance equations.
The Kalman prediction step requires an estimated
posterior covariance step in the previous time step,
which can be estimated through the Kalman update step.
When the dynamic model does not match the linear Gaussian scenario,
the extended Kalman filter (EKF) or
the unscented Kalman filter (UKF)~\cite{julier1997}
covariance prediction equations can be applied to estimate $P$.

\begin{algorithm}[tbh]
\begin{algorithmic}[1]
\STATE Initialization: Draw $\{x_{0}^{i} \}_{i=1}^{N_p}$ from the prior $p_0(x)$. Set $\hat{x}_0$ and $\hat{P}_{0}$ to be the mean and covariance of $p_0(x)$, respectively\;
\STATE Set $\{w_{0}^{i}\}_{i=1}^{N_p} = \frac{1}{N_p}$\;
\FOR {$k = 1$ to $K$}
\FOR {$i = 1,\ldots, N_p$}
\STATE Apply EKF/UKF prediction to estimate $P^i$:
$(x_{k-1}^i, P_{k-1}^i) \rightarrow (m_{k|k-1}^i,P^i)$\;
\STATE Calculate $\bar{\eta}^i = g_k(x_{k-1}^{i},0)$\;
\STATE Propagate particles $\eta_0^i = g_k(x_{k-1}^{i},v_k)$\;
\STATE Set $\eta_1^i = \eta_0^i$ and
 $\theta^i = 1$\;
\STATE Calculate $\bar{\eta}_0^i = g_k(x_{k-1}^i,0)$\;
\ENDFOR
\STATE Set $\lambda = 0$\;
\FOR {$j = 1,\ldots,N_\lambda$}
\STATE Set $\lambda = \lambda + \epsilon_j$\;
\FOR {$i = 1,\ldots, N_p$}
\STATE Set $\bar{\eta}_0 = \bar{\eta}_0^i$ and $P = P^i$\;
\STATE Calculate $A_j^i(\lambda)$ and $b_j^i(\lambda)$
from~\eqref{eq:LEDH_A_nonlinear} and~\eqref{eq:LEDH_b_nonlinear}
with the linearization being performed at $\bar{\eta}^i$\;
\STATE Migrate $\bar{\eta}^i$:
$\bar{\eta}^i = \bar{\eta}^i + \epsilon_j(A_j^i(\lambda)\bar{\eta}^i+b_j^i(\lambda))$\;
\STATE Migrate particles: $\eta_1^{i} = \eta_1^{i} + \epsilon_j(A_j^i(\lambda)\eta_1^i+b_j^i(\lambda))$\;
\STATE Calculate $\theta^i = \theta^i
|\det(I+\epsilon_jA_j^i(\lambda))|$
\ENDFOR
\ENDFOR
\FOR {$i = 1,\ldots, N_p$}
\STATE Set $x_{k}^{i} = \eta_1^{i}$\;
\STATE $w_{k}^{i} =\dfrac{p(x_{k}^{i}|x_{k-1}^{i})p(z_k|x_{k}^{i})\theta^i
}{p(\eta_0^i|x_{k-1}^i)}w_{k-1}^i$\;
\ENDFOR
\FOR {$i = 1,\ldots, N_p$}
\STATE Normalize $w_{k}^{i} = w_{k}^{i}/\sum_{s=1}^{N_p}w_{k}^{s}$\;
\STATE Apply EKF/UKF update:
$(m_{k|k-1}^i, P^i) \rightarrow (m_{k|k}^i,P_{k}^i)$\;
\ENDFOR
\STATE Estimate $\hat{x}_k = \sum_{i=1}^{N_p}w_k^i x_k^{i}$\;
\STATE (Optional) Resample $\{x_{k}^{i}, P_k^i, w_{k}^{i}\}_{i=1}^{N_p}$
to obtain $\{x_{k}^{i}, P_k^i, \frac{1}{N_p}\}_{i=1}^{N_p}$\;
\ENDFOR
\end{algorithmic}
\caption{Particle flow particle filtering (LEDH).}
\label{alg:PF-PF_LEDH}
\end{algorithm}

We now prove that
the function constructed by the discretized particle flow
in Algorithm~\ref{alg:PF-PF_LEDH}, $\eta_1^i = T(\eta_0^i;z_k,x_{k-1}^i)$,
is invertible.
We start with the following lemma.
\begin{lemma}
For any $\lambda\in [0, 1)$,
if $h(\cdot,0)$ is a $C^1$ function
and $\bar{\eta}_{\lambda}^i$ is bounded,
$\rho(A^i(\lambda))$ of $A^i(\lambda)$
defined by~\eqref{eq:LEDH_A_nonlinear} is upper-bounded.
Here $\rho(\cdot)$ denotes the spectral radius.
\label{lem:A_bound}
\end{lemma}
\begin{IEEEproof}
Denote the largest eigenvalue of $P$
by $\bar{p}$, and the smallest eigenvalue of $(\lambda H^i(\lambda)^T P H^i(\lambda) + R)$ by
$\underline{r}(\lambda)$. Since $P$ is positive definite, 
$\bar{p} > 0$. Since $R$ is positive definite,
for any non-zero $\mu \in \mathbb{R}^d$, 
\begin{align}
&\mu^T\left(\lambda H^i(\lambda)^T P H^i(\lambda) + R\right)\mu \nonumber \\
& \quad \quad \quad \quad = \lambda \left(H^i(\lambda)\mu\right)^T P \left(H^i(\lambda)\mu\right) + \mu^T R \mu \nonumber\\
&\quad \quad \quad \quad > 0\,.
\end{align}
Thus, $(\lambda H^i(\lambda)^T P H^i(\lambda) + R)$ is positive definite.
So, $\underline{r}(\lambda)>0$.

Denote the operator norm induced by the Euclidean norm by $||\cdot||$. Since $P$ and $\left(\lambda H^i(\lambda)^T P H^i(\lambda) + R\right)^{-1}$ are both positive semi-definite, $||P||$
and $||(\lambda H^i(\lambda)^T P H^i(\lambda) + R)^{-1}||$ are
equal to the spectral radius, and we have
\begin{align}
||P|| = \bar{p}\,,
\end{align}
\begin{align}
||(\lambda H^i(\lambda)^T P H^i(\lambda) + R)^{-1}||
&= \frac{1}{\underline{r}(\lambda)}\,.
\end{align}

We also have
\begin{align}
||H^i(\lambda)|| &=\sqrt{\rho_{\max}(H^i(\lambda)^T H^i(\lambda))}\nonumber\\
&\leq \sqrt{\mbox{Tr}(H^i(\lambda)^T H^i(\lambda))}\nonumber\\
&=||H^i(\lambda)||_F\,,
\end{align}
where $\mbox{Tr}(\cdot)$ denotes
the trace of a matrix, and $||\cdot||_F$
is the Frobenius norm. Similarly, $||H^i(\lambda)^T|| \leq ||H^i(\lambda)||_F$.

Since $h(\cdot)$ is a $C^1$ function, $H^i(\lambda)$ is continuous on
$\bar{\eta}_{\lambda}^i \in \mathbb{R}^d$. Since $\bar{\eta}_{\lambda}^i$ is bounded on
$\lambda \in [0, 1]$,
$H^i(\lambda)$ is bounded on $\lambda \in [0, 1]$.
Thus, there exists an $\bar{h} > 0$ such that $||H^i(\lambda)||_F \leq \bar{h}$ for any $\lambda \in [0, 1]$.
%

For the square matrix $A^i(\lambda)$,
its spectral radius is upper-bounded by
its operator norm.
Thus, by the sub-multiplicativity of the operator norm,
\begin{align}
\rho(A^i(\lambda))&\leq||A^i(\lambda)|| \nonumber\\
=& ||-\dfrac{1}{2}PH^i(\lambda)^T(\lambda H^i(\lambda)PH^i(\lambda)^T+R)^{-1}H^i(\lambda)||\nonumber\\
\leq & \dfrac{1}{2}||P|| \cdot ||H^i(\lambda)^T|| \cdot ||(\lambda H^i(\lambda)^T P H^i(\lambda) + R)^{-1}||\nonumber\\
& \cdot ||H^i(\lambda)||\nonumber\\
\leq & \dfrac{\bar{p}\bar{h}^2}{2\underline{r}(\lambda)}\,.
\label{eq:A_bound}
\end{align} 
\end{IEEEproof}

We now prove that with a sufficiently small
step size, the mapping defined by~\eqref{eq:EDH_A_nonlinear},~\eqref{eq:EDH_b_nonlinear}
and~\eqref{eq:discrete_update} is invertible.

\begin{lemma}
For any $\lambda_j \in [0, 1)$,
if $h(\cdot,0)$ is a $C^1$ function and
$\bar{\eta}_{\lambda_j}^i$ is bounded,
$f_{\lambda_j}^i$ defined by~\eqref{eq:LEDH_A_nonlinear},
\eqref{eq:LEDH_b_nonlinear} and~\eqref{eq:discrete_update_LEDH}
is invertible, if $\epsilon_j$ is sufficiently small, specifically $\epsilon_j <
\dfrac{2\underline{r}(\lambda_j)}{\bar{p}\bar{h}^2}$.
\label{lem:LEDH_nonlinear}
\end{lemma}

\begin{IEEEproof}
For any $i \in \{1,\ldots,N_p\}$
and $\lambda_j \in [0, 1)$,
consider two values of
$\eta_{\lambda_j}^i$, $\eta \neq \eta'$,
Since $A^i(\lambda_j)$ and $b^i(\lambda_j)$ are the same
for $\eta \neq \eta'$,
from~\eqref{eq:LEDH_mapping} and~\eqref{eq:discrete_update_LEDH},
\begin{align}
f_{\lambda_j}^i(\eta) = \eta
+ \epsilon_j(A^i(\lambda_j)\eta+b^i(\lambda_j))\,,\\
f_{\lambda_j}^i(\eta') = \eta'
+ \epsilon_j (A^i(\lambda_j)\eta'+b^i(\lambda_j))\,.
\end{align}
If $f_{\lambda_j}^i(\eta)
= f_{\lambda_j}^i(\eta')$, then
\begin{align}
\eta - \eta' = -\epsilon_j A^i(\lambda_j)
(\eta - \eta')\,.
\label{eq:LEDH_equal_nonlinear}
\end{align}

Equation~\eqref{eq:LEDH_equal_nonlinear} holds only if
$\eta - \eta'$ is an eigenvector
of $A^i(\lambda_j)$ and its corresponding eigenvalue $\psi$
satisfies $\epsilon_j \psi= -1$.
From Lemma~\ref{lem:A_bound},
$\rho_{\max}(A^i(\lambda_j))$ is upper-bounded for $\lambda_j \in [0, 1)$.
Thus $|\psi| \leq \dfrac{\bar{p}\bar{h}^2}{2\underline{r}(\lambda_j)}$.
If we choose $\epsilon_j <
\dfrac{2\underline{r}(\lambda_j)}{\bar{p}\bar{h}^2}$ then $\epsilon_j|\psi| < 1$ and $f_{\lambda_j}^i(\eta)
\neq f_{\lambda_j}^i(\eta')$, implying that $f_{\lambda_j}^i$
is injective.

For this choice of $\epsilon_j$, the equality~\eqref{eq:LEDH_equal_nonlinear}
does not hold unless $\eta \neq \eta'$. Thus
\begin{align}
(I+\epsilon_j A^i(\lambda_j))
(\eta - \eta')=0\,,
\label{eq:LEDH_equal_support}
\end{align}
only holds for $\eta = \eta'$, demonstrating that
Null$(I + \epsilon_j A^i(\lambda_j)) = \{0\}$.
Hence,
\begin{align}
\mbox{dim(range}(I + \epsilon_j A^i(\lambda_j)))
&= d - \mbox{dim(Null}(I + \epsilon_j A^i(\lambda_j)))\nonumber\\
&= d
\end{align}
Since range$(I + \epsilon_j A^i(\lambda_j))$ is a subspace
of $\mathbb{R}^{d}$,
\begin{align}
\mbox{range}(I + \epsilon_j A^i(\lambda_j)) = \mathbb{R}^{d}\,.
\label{eq:full_rank}
\end{align}
Thus, $(I + \epsilon_j A^i(\lambda_j))$
has full rank and the mapping $f_{\lambda_{j}}^i(\eta_{\lambda_{j}}^i)
= (I + \epsilon_j A^i(\lambda_j))\eta_{\lambda_j}^i+b^i(\lambda_j)$
is surjective.

Thus for the specified choice of $\epsilon_j$, $f_{\lambda_j}^i$ is both
injective and surjective, and hence invertible.
\end{IEEEproof}

Now we can establish the following theorem:
\begin{theorem}
If $h(\cdot,0)$ is a $C^1$ function, $\eta_1^i = T(\eta_0^i;z_k,x_{k-1}^i)$
in the particle flow particle filter with the LEDH flow
defines an invertible mapping,
if $\epsilon_j < \dfrac{2\underline{r}(\lambda_j)}{\bar{p}\bar{h}^2}$ for $j\in \{1,\ldots,N_\lambda\}$ .
\label{tem:LEDH_mapping}
\end{theorem}

\begin{IEEEproof}
The theorem follows directly from Lemma~\ref{lem:LEDH_nonlinear} if
$\bar{\eta}^i_{\lambda_{j}}$ is bounded for $j\in \{1,\ldots,N_\lambda\}$.
We prove this by induction. 
For $j=1$, $\bar{\eta}^i_{\lambda_{1}} = \bar{\eta}^i_0$ is bounded
as it is generated by propagating the
sample mean of particles in the previous
time step using $g(\cdot,0)$ which is a bounded function.

For $j \in \{1,\ldots,N_{\lambda}-1\}$ assume it is true that
$\bar{\eta}^i_{\lambda_j}$ is bounded. For the specified choice of $\epsilon_j$,
from Lemma~\ref{lem:LEDH_nonlinear},
$f_{\lambda_j}^i$ is invertible. Thus,
$\bar{\eta}_{\lambda_{j+1}}^i = f_{\lambda_j}^i(\bar{\eta}^i_{\lambda_j})$ is bounded.

By induction, $\bar{\eta}^i_{\lambda_{j}}$ is bounded for all $j\in \{1,\ldots,N_\lambda\}$.
This implies from Lemma~\ref{lem:LEDH_nonlinear} that
$f_{\lambda_j}^i$ is invertible for $j\in
\{1,\ldots,N_\lambda\}$. Since the deterministic mapping
\begin{align}
\eta_1^i &= T(\eta_0^i;z_k,x_{k-1}^i) \nonumber\\
&=f_{\lambda_{{N_\lambda}}}^i(\ldots f_{\lambda_1}^i(\eta_0^i))
\end{align}
is a chain of invertible mappings,
$T(\eta_0^i)$ is an invertible mapping.
\end{IEEEproof}

We now prove that $0<|\dot{T}(\eta_0^i;z_k,x_{k-1}^i)|<\infty$ for the specified
choice of $\epsilon_j$, which shows
that the importance weight is finite for all $\eta_1^i$ generated by
applying the constructed mapping $T(\eta_0^i)$.
\begin{lemma} 
If $\epsilon_j < \dfrac{2\underline{r}(\lambda_j)}{\bar{p}\bar{h}^2}$
for $j\in \{1,\ldots,N_\lambda\}$, then $0<|\dot{T}(\eta_0^i;z_k,x_{k-1}^i)|<\infty$.
\label{lem_mapbound}
\end{lemma}

\begin{IEEEproof}
From~\eqref{eq:full_rank},
we see that $(I+\epsilon_jA^i_j(\lambda))$
is invertible, $\forall j \in \{1,\ldots,N_\lambda\}$.
Thus, $\det(I+\epsilon_jA^i_j(\lambda))\neq 0, \forall j \in \{1,\ldots,N_\lambda\}$.
From~\eqref{eq:jac_det_LEDH},
\begin{equation}
|\dot{T}(\eta_0^i;z_k,x_{k-1}^i)| =\prod_{j=1}^{N_\lambda}|\det(I+\epsilon_jA_j^i(\lambda))|>0\,.
\label{eq:det_lower_bound}
\end{equation}
We denote the $m$-th column of
$(I+\epsilon_jA_j^i(\lambda))$
by $u_j^m \in \mathbb{R}^d$, for $m\in \{1,\ldots,d\}$.
From Hadamard's inequality,
\begin{align}
\det(I+\epsilon_jA_j^i(\lambda))
\leq\prod_{m=1}^{d} ||u_j^m||_2\,,
\end{align}
where $||\cdot||_2$ is the Euclidean norm.
Thus,
\begin{align}
|\dot{T}(\eta_0^i;z_k,x_{k-1}^i)| &=\prod_{j=1}^{N_\lambda}|\det(I+\epsilon_jA_j^i(\lambda))|\nonumber\\
&\leq\prod_{j=1}^{N_\lambda}\prod_{m=1}^{d} ||u_j^m||_2<\infty\,.
\label{eq:det_upper_bound}
\end{align}
\end{IEEEproof}

\subsection{Particle Flow Particle Filtering with the EDH flow}
\label{sec:PFPF_EDH}

The particle flow particle filter algorithm based on the
EDH flow with the invertible mapping property is presented in Algorithm~\ref{alg:PF-PF_EDH}.
The PF-PF (EDH) is much more computationally efficient
that the PF-PF (LEDH) by using common flow parameters $A_j(\lambda)$ and $b_j(\lambda)$ to perform flows
for different particles. But this leads to statistical
correlations between particles and the effect of the dependence can be challenging to characterize. Hence,
the PF-PF (EDH) reduces the computational cost
with a compromise on convergence properties
of standard particle filters~\cite{crisan2002}.

The mapping
defined by~\eqref{eq:EDH_A_nonlinear}
and~\eqref{eq:EDH_b_nonlinear} is
the same as that defined
by~\eqref{eq:LEDH_A_nonlinear} and~\eqref{eq:LEDH_b_nonlinear},
if we replace $\bar{\eta}^i$ by $\bar{\eta}$ in Algorithm~\ref{alg:PF-PF_LEDH}.
From Theorem~\ref{tem:LEDH_mapping},
this defines an invertible mapping between $\eta_0^i$
and $\eta_1^i$ for any $i$.
Thus, the weight update equation~\eqref{eq:weight_update} still holds for
Algorithm~\ref{alg:PF-PF_EDH}.

\begin{algorithm}[tbh]
\begin{algorithmic}[1]
\STATE Initialization: Draw $\{x_{0}^{i} \}_{i=1}^{N_p}$ from $p_0(x)$.
Set $\hat{x}_0$ and $\hat{P}_{0}$ to be the mean and covariance of $p_0(x)$, respectively\;
\STATE Set $\{w_{0}^{i}\}_{i=1}^{N_p} = \frac{1}{N_p}$\;
\FOR {k = 1 to T}
\STATE Apply EKF/UKF prediction to estimate $P$:
$(\hat{x}_{k-1}, P_{k-1}) \rightarrow (m_{k|k-1},P)$\;
\FOR {$i = 1,\ldots, N_p$}
\STATE Propagate particles $\eta_0^i = g_k(x_{k-1}^{i},v_k)$\;
\STATE Set $\eta_1^i = \eta_0^i$\;
\ENDFOR
\STATE Calculate $\bar{\eta}_0 = g_k(\hat{x}_{k-1},0)$ and set $\bar{\eta} = \bar{\eta}_0$\;
\STATE Set $\lambda = 0$\;
\FOR {$j = 1,\ldots,N_\lambda$}
\STATE Set $\lambda = \lambda + \epsilon_j$\;
\STATE Calculate $A_j(\lambda)$ and $b_j(\lambda)$
from Equation~\eqref{eq:EDH_A_nonlinear}
and ~\eqref{eq:EDH_b_nonlinear}
with the linearization being performed at $\bar{\eta}$\;
\STATE Migrate $\bar{\eta}$:
$\bar{\eta} = \bar{\eta} + \epsilon_j( A_j(\lambda)\bar{\eta}+b_j(\lambda))$\;
\FOR {$i = 1,\ldots, N_p$}
\STATE Migrate particles: $\eta_1^{i} = \eta_1^{i} + \epsilon_j(A_j(\lambda)\eta_1^i+b_j(\lambda))$\;
\ENDFOR
\ENDFOR
\FOR {$i = 1,\ldots, N_p$}
\STATE Set $x_{k}^{i} = \eta_1^{i}$\;
\STATE $w_{k}^{i} =\dfrac{p(x_{k}^{i}|x_{k-1}^{i})p(z_k|x_{k}^{i})}{p(\eta_0^i|x_{k-1}^i)}w_{k-1}^i$\;
\ENDFOR
\FOR {$i = 1,\ldots, N_p$}
\STATE Normalize $w_{k}^{i} = w_{k}^{i}/\sum_{s=1}^{N_p}w_{k}^{s}$\;
\ENDFOR
\STATE Apply EKF/UKF update:
$(m_{k|k-1}, P) \rightarrow (m_{k|k},P_{k})$\;
\STATE Estimate $\hat{x}_k = \sum_{i=1}^{N_p}w_k^i x_k^{i}$\;
\STATE (Optional) Resample $\{x_{k}^{i}, w_{k}^{i}\}_{i=1}^{N_p}$
to obtain $\{x_{k}^{i}, \frac{1}{N_p}\}_{i=1}^{N_p}$\;
\ENDFOR
\end{algorithmic}
\caption{Particle flow particle filtering (EDH).}
\label{alg:PF-PF_EDH}
\end{algorithm}

For the PF-PF (EDH), we do not need to evaluate the Jacobian
determinant.
Based on the migration of particles shown
in Line 16 of Algorithm~\ref{alg:PF-PF_EDH},
\begin{align}
|\dot{T}(\eta_0^i;z_k,x_{k-1}^i)| &=
|\det(\dfrac{d\eta_1^i}{d\eta_0^i})|\nonumber\\
&=|\det(\dfrac{d[\prod_{j=1}^{N_\lambda}(I+\epsilon_jA_j(\lambda))]\eta_0^i}{d\eta_0^i})|\nonumber\\
&=\prod_{j=1}^{N_\lambda}|\det(I+\epsilon_jA_j(\lambda))|\,,
\label{eq:jac_det_EDH}
\end{align}
where $\det(\cdot)$ denotes the determinant.
From~\eqref{eq:jac_det_EDH}, we see that for the PF-PF
(EDH), the determinant is the same for different particles, i.e.,
$|\dot{T}(\eta_0^i;z_k,x_{k-1}^i)| = |\dot{T}(\eta_0^{i'};z_k,x_{k-1}^{i'})|$ for $i\neq
i^{'}$.

We can show that $0<|\dot{T}(\eta_0^i;z_k,x_{k-1}^i)|<\infty$
using the same argument as in Lemma~\ref{lem_mapbound}.
Thus, the weight update expression for
the PF-PF (EDH) is
\begin{align}
  w_{k}^{i} \propto
  \dfrac{p(\eta_1^{i}|x_{k-1}^{i})p(z_k|\eta_1^{i})}{p(\eta_0^i|x_{k-1}^i)}w_{k-1}^i\,.
  \label{eq:weight_update_PFPF_EDH} 
\end{align}

\subsection{Implementation and Complexity}

Several numerical integration schemes are proposed and discussed in~\cite{daum2013a,daum2014b,khan2015}.
For algorithms involving particle flows,
we adopt the $N_\lambda = 29$ exponentially spaced step sizes recommended in~\cite{daum2013a}.
The constant ratio between step sizes is 1.2, i.e.
$q = \frac{\epsilon_j}{\epsilon_{j-1}} = 1.2$,
for $j = 2, 3, \ldots, N_\lambda$. The initial step size
$\epsilon_1 = \frac{1-q}{1-q^{N_\lambda}} \approx 0.001$. 

With the cost of increased
computation, the eigenvalues
of $A^i(\lambda)$ or $A(\lambda)$ can be evaluated and an adaptive step size used to 
ensure that the invertible mapping property is satisfied. In practice it is very unlikely
that any of the pre-defined step sizes satisfies $\epsilon_j\psi = -1$ where $\psi$ is any eigenvalue of $A^i(\lambda)$ or $A(\lambda)$ .
Denote by $\rho_{\max}(A)$ the largest magnitude of any eigenvalue of a matrix $A$. We have checked that the step-size choice described above leads to $\epsilon_j$ values that are always smaller than $\frac{1}{\rho_{\max}(A^i(\lambda))}$ or $\frac{1}{\rho_{\max}(A(\lambda))}$ in the simulation scenarios examined in Section~\ref{sec:simulation}. 

For the PF-PF (LEDH), 
the most computationally demanding part of the algorithm is the inverse operation in calculating $A^i(\lambda)$ and $b^i(\lambda)$.
Since individual flow parameters are calculated for each particle,
the computational complexity of the matrix inverse operations is $O(N_p S^{3})$ (recall that $S$ is the measurement dimension).
There is an additional overhead in calculating the determinant in line
19 of Algorithm~\ref{alg:PF-PF_LEDH}, but this is small
compared to the inverse operations. Once the $\theta_i$ values in line
19 have been computed, the
complexity of the weight update in line 24
is $O(N_p)$. The computational cost of the weight update is much lower than that of the GPFIS,
which involves calculating matrix square roots and repeatedly solving
the Sylvester equation.

For the PF-PF (EDH) introduced in Section~\ref{sec:PFPF_EDH},
the most computationally intensive part of the flow is again the inversion
operation in Equation~\eqref{eq:EDH_A_nonlinear}
and~\eqref{eq:EDH_b_nonlinear}, which has a computation complexity of
$O(S^{3})$. Since the calculation of the flow parameters is only
performed at $\bar{\eta}$, the computational complexity of the inverse
operation does not depend on the number of particles $N_p$. 
The weight update does not depend on the number of intermediate flow update steps
$N_{\lambda}$ as no determinant needs to be calculated
to update the importance weight. The computational cost of the weight calculation is
usually negligible compared to that of the flow; an exception is when
the prior probability $p(\eta_1^{i}|x_{k-1}^i)$ is difficult to evaluate.

One possible concern with the proposed implementation is that flow parameters are calculated using the auxiliary flows by linearizing at $\bar{\eta}^i$ (LEDH) or $\bar{\eta}$ (EDH), as opposed to linearizing at the actual particle values $\{\eta^i\}$. We note that the linearization in the EDH is already performed at the mean or median of the particle cloud, so it is unlikely that using $\bar{\eta}$ introduces additional error for the PF-PF (EDH). For the PF-PF (LEDH), however, there is the potential for additional error beyond the linearization due to this mismatch. We have compared the original LEDH filter with an LEDH filter that uses the auxiliary flows for flow parameter calculation using examples from Section~\ref{sec:simulation} (for conciseness, these results are not shown in the paper). The constructed flows and the filtering performance were very similar, suggesting that the use of auxiliary flows introduces minimal error for these scenarios. This issue does warrant further exploration, and ideally a filter can be designed such that weight updates can be performed with small computational cost and the flows are calculated using the actual particle values. 

\section{Simulations and Results}
\label{sec:simulation}

We explore the performance of the PF-PF algorithms
in two challenging simulation setups.
The first is a multi-target acoustic
tracking scenario with small
measurement noise. The second is a high
dimensional filtering problem in which the state evolves according to a multivariate Generalized Hyperbolic (GH)
skewed-t distribution and the observations are count data derived via a Poisson process.  
Both scenarios lead to severe
particle degeneracy for bootstrap particle
filters due to either the highly informative
measurements or the high dimensionality.
In addition, we compare the performance of the PF-PF
with the optimal filter in a simple linear Gaussian filtering example.
Matlab code implementing the simulation experiments is available~\footnote{http://networks.ece.mcgill.ca/sites/default/files/PFPF.zip}.

\subsection{Multi-target acoustic tracking}
\label{sec:acoustic}

\subsubsection{Simulation setup}

We constructed a multi-target tracking scenario
with a relatively large state space and
highly informative measurements, based on
the simulation setup proposed in~\cite{hlinka2011}.
There are $C = 4$ targets moving independently in a region
of size of $40$ m $\times 40$ m. Each follows a
constant velocity model
$x_{k}^{(c)} = F x_{k-1}^{(c)} + v_k^{(c)}$,
where $x_k^{(c)} = [\mathrm{x}_k^{(c)},\mathrm{y}_k^{(c)},
\dot{\mathrm{x}}_k^{(c)},\dot{\mathrm{y}}_k^{(c)}]$ are
the position and velocity components of the $c$-th
target. $F = \begin{bmatrix}
1 & 0 & 1 & 0 \\
0 & 1 & 0 & 1 \\
0 & 0 & 1 & 0 \\
0 & 0 & 0 & 1
\end{bmatrix}$ is the state transition matrix.
$v_k^{(c)} \sim N(0,V)$ is the process noise.

\begin{figure}[!ht]
 \centering
 \includegraphics[width=.48\textwidth]{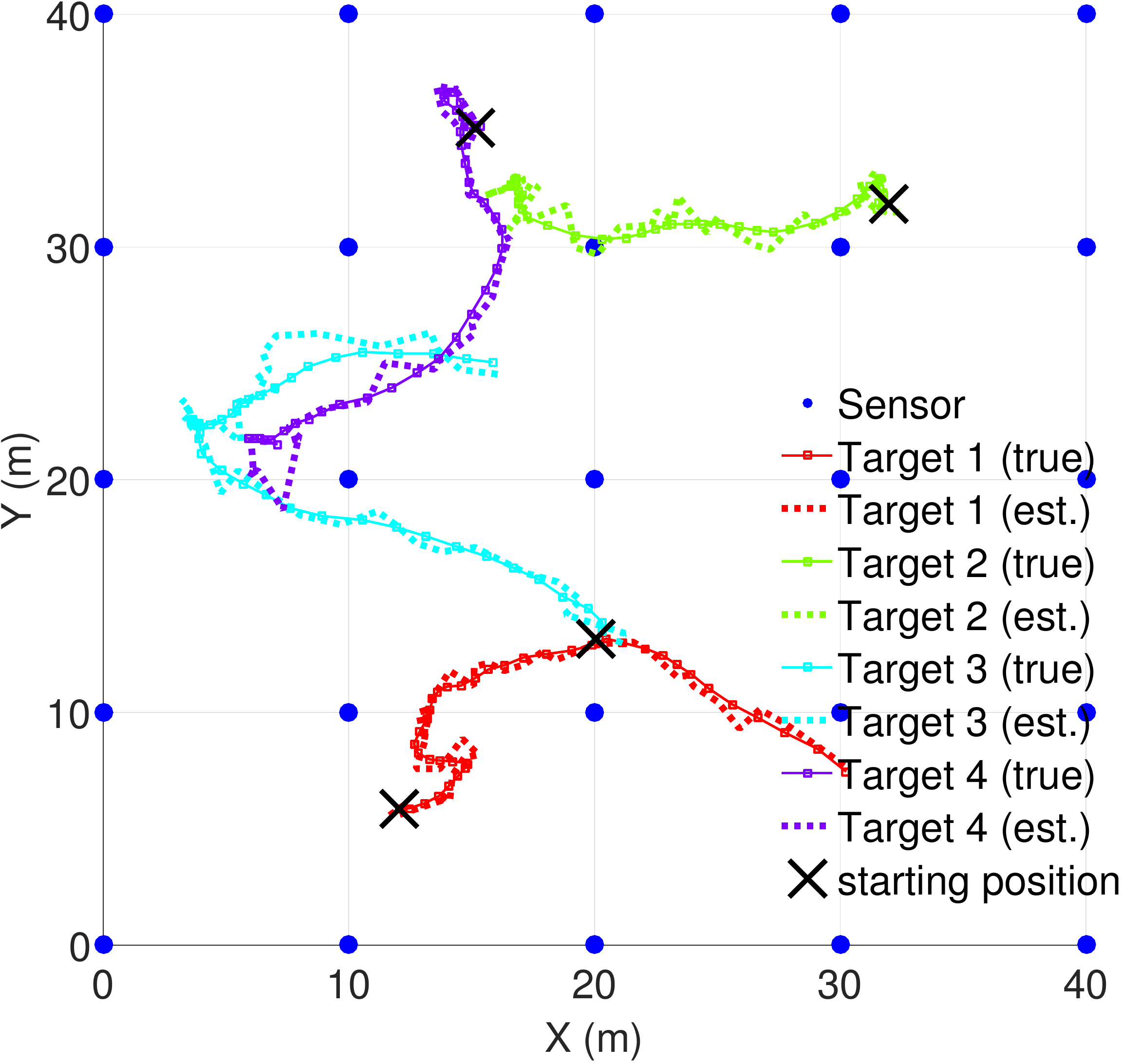}
 \caption{\label{fig:estimated_trajectory}
 One example of estimated trajectories using PF-PF (LEDH).
 The crosses mark the starting positions of the four targets and
 the solid lines show their true trajectories. Dotted lines
 indicate the estimated trajectories.}
\end{figure}

At each time step, all targets emit sounds of amplitude $\Psi$.
Attenuated sounds are measured by all sensors. Each sensor only records the sum of amplitudes.
Thus, the measurement function for the $s$-th sensor located at $R^s$
is additive:
\begin{align}
\bar{z}^s(x_k) = \sum_{c=1}^{C}\dfrac{\Psi}{||(\mathrm{x}_k^{(c)},\mathrm{y}_k^{(c)})^T-R^s||_2+d_0}\,,
\end{align}
where $||\cdot||_2$ is the Euclidean norm,
$d_0 = 0.1$ and $\Psi = 10$.
There are $N_s = 25$ sensors located at grid 
intersections within the tracking area,
as shown in Figure~\ref{fig:estimated_trajectory}.
The measurements are perturbed by Gaussian
noise, i.e., the noisy measurement $z_k^s$ from the $s$-th sensor
is drawn from $N(\bar{z}^s(x_k),\sigma_w^2)$. $\sigma_w^2$ is set to $0.01$.
This leads to very informative measurements.

The initial target states are
$[12,6,0.001,0.001]^T$, $[32,32,-0.001,-0.005]^T$, $[20,13,-0.1,0.01]^T$
and \newline $[15,35,0.002,0.002]^T$. 100 random trajectories are simulated
using a constant velocity model with the process covariance matrix
$\dfrac{1}{20}\begin{bmatrix}
1/3 & 0 & 0.5 & 0 \\
0 & 1/3 & 0 & 0.5 \\
0.5 & 0 & 1 & 0 \\
0 & 0.5 & 0 & 1
\end{bmatrix}$.
One set of measurements is generated
for each trajectory. We run each algorithm 5 times on each measurement set. Each execution starts with a different initial distribution. We implement the simulation using Matlab.

\subsubsection{Parameter values for the filtering algorithms}
\label{sec:param_acoustic}

The mean of the initial distributions for the filtering algorithms
is sampled from a Gaussian centered at the true initial states.
The standard deviation is $10$ for positions and $1$ for velocities.
If the initial mean is outside of the tracking area,
we reject and resample it.
The covariance of the process noise for the filters is set as $\begin{bmatrix}
3 & 0 & 0.1 & 0 \\
0 & 3 & 0 & 0.1 \\
0.1 & 0 & 0.03 & 0 \\
0 & 0.1 & 0 & 0.03
\end{bmatrix}$.
The entries are larger than those used to generate the target trajectories,
because we assume that there is more uncertainty about the model during tracking.
Resampling is performed when the effective sample size (ESS) is less than $\frac{N_p}{2}$.
The ESS at time step $k$ is estimated as $\tfrac{1}{\sum_{i=1}^{N_p} (w_k^i)^2}$ after weight normalization.
$N_p = 500$ particles are used in all Monte Carlo-based algorithms except for the BPF. 
The EKF is used to estimate the predictive covariance
needed to calculate the flow parameters in Equation~\eqref{eq:LEDH_A_nonlinear} and \eqref{eq:LEDH_b_nonlinear}.
The EDH and LEDH filter implementations adopt the redraw strategy
in~\cite{ding2012} at the beginning of each time step.
$2d+1$ sigma points are generated for each particle in the UPF,
where $d = 16$ is the state dimension.
We use the ensemble square root filter (ESRF)~\cite{evensen2004}, a popular implementation of the ensemble Kalman filter (EnKF), to construct the transport map
for the GSMC~\cite{reich2013}.
We set the diffusion term for the GPFIS algorithm to 0,
as suggested in~\cite{bunch2016}.

\subsubsection{Experimental results}

For this simulation scenario, we compare the PF-PF (LEDH) and PF-PF (EDH) 
algorithms proposed in this paper with the GPFIS 
algorithm~\cite{bunch2016}, the BPF~\cite{gordon1993}, and the EDH and 
LEDH particle flow algorithms~\cite{daum2010a,ding2012},
the GSMC~\cite{reich2013} and various Kalman-type filters~\cite{evensen2004,julier1997}.
For this example, we do not compare with the SmHMC 
algorithm~\cite{septier2016}, because the target marginal posterior 
distribution is not log-concave, so the negative Hessian is not globally 
positive-definite, rendering implementation of SmHMC more challenging. We 
also do not compare with the block particle filter~\cite{rebeschini2015}, 
because identifying a suitable partitioning of the state space is 
difficult, since many state variables contribute to each measurement.

The error metric we use in this multi-target tracking scenario
with a fixed number of targets is the optimal mass transfer (OMAT)
metric~\cite{schuhmacher2008}. The OMAT metric
$d_p(X,\hat{X})$ between two arbitrary sets
$X = \{x_1,x_2,\ldots,x_C\}$ and
$\hat{X} = \{\hat{x}_1,\hat{x}_2,\ldots,\hat{x}_C\}$
is defined as
\begin{align}
d_p(X,\hat{X}) = (\frac{1}{C}\min_{\pi\in\Pi}\sum_{c=1}^{C}d(x_c,\hat{x}_{\pi(c)})^p)^{1/p}\;
\end{align}
where the scalar $p$ is a fixed parameter,
$\Pi$ is the set of possible permutations of
$\{1,2,\ldots,C\}$,
and $d(x,\hat{x})$ is the Euclidean distance
between $x$ and $\hat{x}$. We set $p$ to 1,
so the OMAT metric assigns targets using
the permutation that minimizes the Euclidean
distance to the true target positions.

\begin{figure}[htbp]
 \centering
 \includegraphics[width=.48\textwidth]{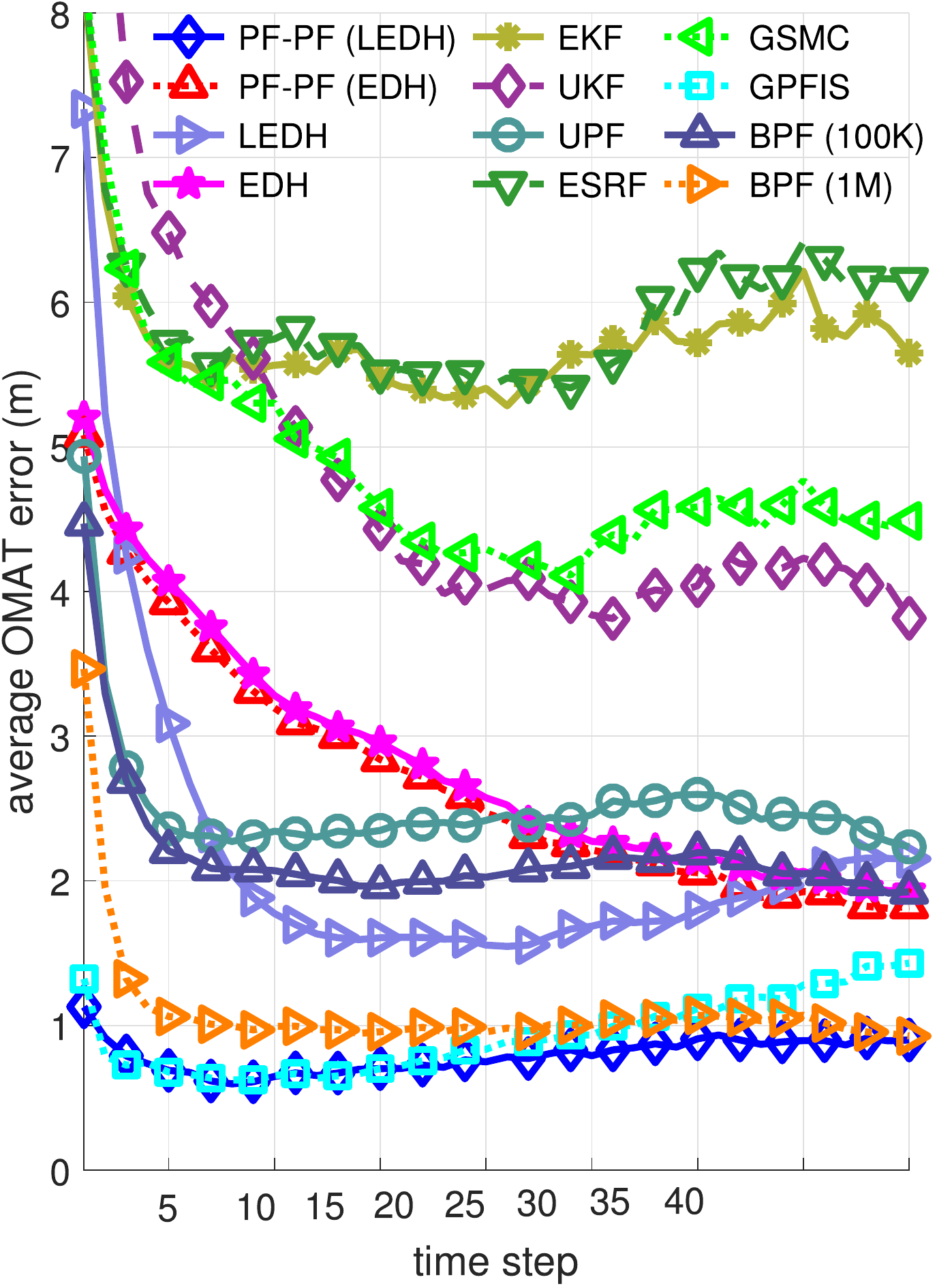}
 \caption{\label{fig:error_acoustic_average}
 Average OMAT errors at each time step
 in the multi-target acoustic tracking example.}
\end{figure}

Figure~\ref{fig:error_acoustic_average} shows the average OMAT metric
at each time step for the various tracking algorithms we
compare. The PF-PF (LEDH) exhibits the smallest average
tracking error, and reduces the average OMAT
below 2 meters with just one time step.
A sample of the estimated trajectories is shown in 
Figure~\ref{fig:estimated_trajectory}.
The PF-PF (LEDH) has much better performance than the LEDH flow algorithm
which it uses to generate the proposal distribution.
This demonstrates the benefits brought by the importance sampling step in 
the PF-PF (LEDH).
We also observe that the invertible particle flow procedure we design
based on the LEDH flow has similar performance to the LEDH filter.
The EDH filter leads to much larger average tracking errors than the LEDH filter.
The PF-PF (EDH) is much less accurate than the PF-PF (LEDH),
indicating that the proposal distribution constructed
using the EDH flow does not provide a good match to the posterior 
distribution. When the measurement function $h$ varies significantly over 
the state space, it is important to perform local linearization and apply 
different mapping functions to different particles.  

\begin{figure}[htbp]
 \centering
 \includegraphics[width=.48\textwidth]{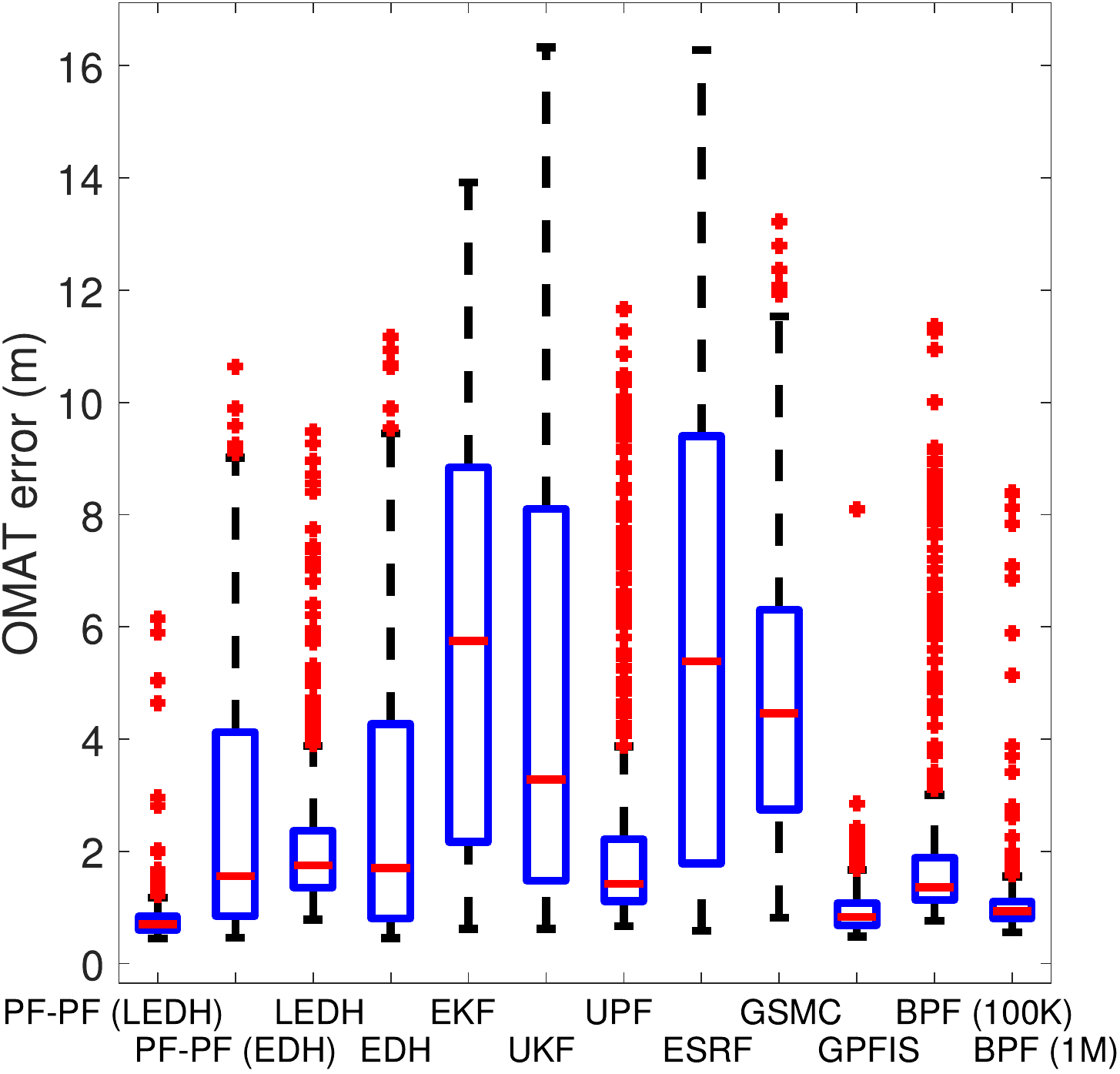}
 \caption{\label{fig:boxplot_OMAT}
 Boxplots of average OMAT errors in
 the multi-target acoustic tracking example.}
\end{figure}

The GPFIS also has impressive tracking performance in the first 20 time steps.
However, the estimation error increases in
later time steps, possibly due to the fact
that the weight update is approximate,
unless the integration step size goes to 0.
However, this is not computationally feasible, since even with 29 discrete time steps, GPFIS
is the most computationally expensive algorithm, as shown in Table~\ref{tab:time_error_acoustic}.
The BPF with 1 million particles has the second
smallest average error in the later time steps,
significantly smaller than BPF with $10^5$ particles.
All tested variants of Kalman-type filters, including the EKF,
the ESRF, and the UKF, exhibited very poor average tracking performance,
probably because of the strong non-linearity of the measurement function.
The UPF has smaller average estimation error than the UKF,
but the average ESS is still small. The GSMC method has relatively poor performance, due to the fact that the ESRF it uses to construct
the transport map does not generate very good proposal distributions.
The boxplots of average (over time) OMAT (Figure~\ref{fig:boxplot_OMAT})
present similar performance relations.
The PF-PF (LEDH) has the smallest median error as well as the first and third quartiles.
There are also far fewer outliers, which are possible indicators of lost 
tracks, than for most of the other algorithms.

\begin{table}[htbp]
\centering
\tabcolsep=0.03cm
\caption{Average OMAT metrics, ESS and execution time per step.
Results are produced with an Intel i7-4770K 3.50GHz CPU and 32GB RAM.}
\label{tab:time_error_acoustic}
\renewcommand{\arraystretch}{1.4}
{
\begin{tabular}{|c|c|c|c|c|}
	\hline
	Algorithm & Particle num.  &
    Avg. OMAT (m) & Avg. ESS & Exec. time (s) \\\hline
    PF-PF (LEDH) & 500 & 0.79 & 45 & 0.9\\ \hline
    PF-PF (EDH) & 500 & 2.71 & 34 & 0.01\\ \hline
    LEDH & 500 & 2.19 & N/A & 0.8\\ \hline
    EDH & 500 & 2.81 & N/A & 0.01\\ \hline
    EKF & N/A & 5.74 & N/A & 0.00003\\\hline
    UKF & N/A & 4.91 & N/A & 0.005\\\hline
    UPF & 500 & 2.51 & 1.48 & 2.0\\\hline
    ESRF & 500 & 5.90 & N/A & 0.01\\\hline
    GSMC & $500$ & 4.87 & 3.5 & 1.6\\\hline
    GPFIS & 500 & 0.93 & 30 & 66.8\\\hline
    BPF & $10^5$ & 2.18 & 2.1 & 0.3\\\hline
    BPF & $10^6$ & 1.10 & 6.3 & 3.0\\\hline
\end{tabular}}
\end{table}

\begin{figure}[htbp]
 \centering
 \includegraphics[width=.48\textwidth]{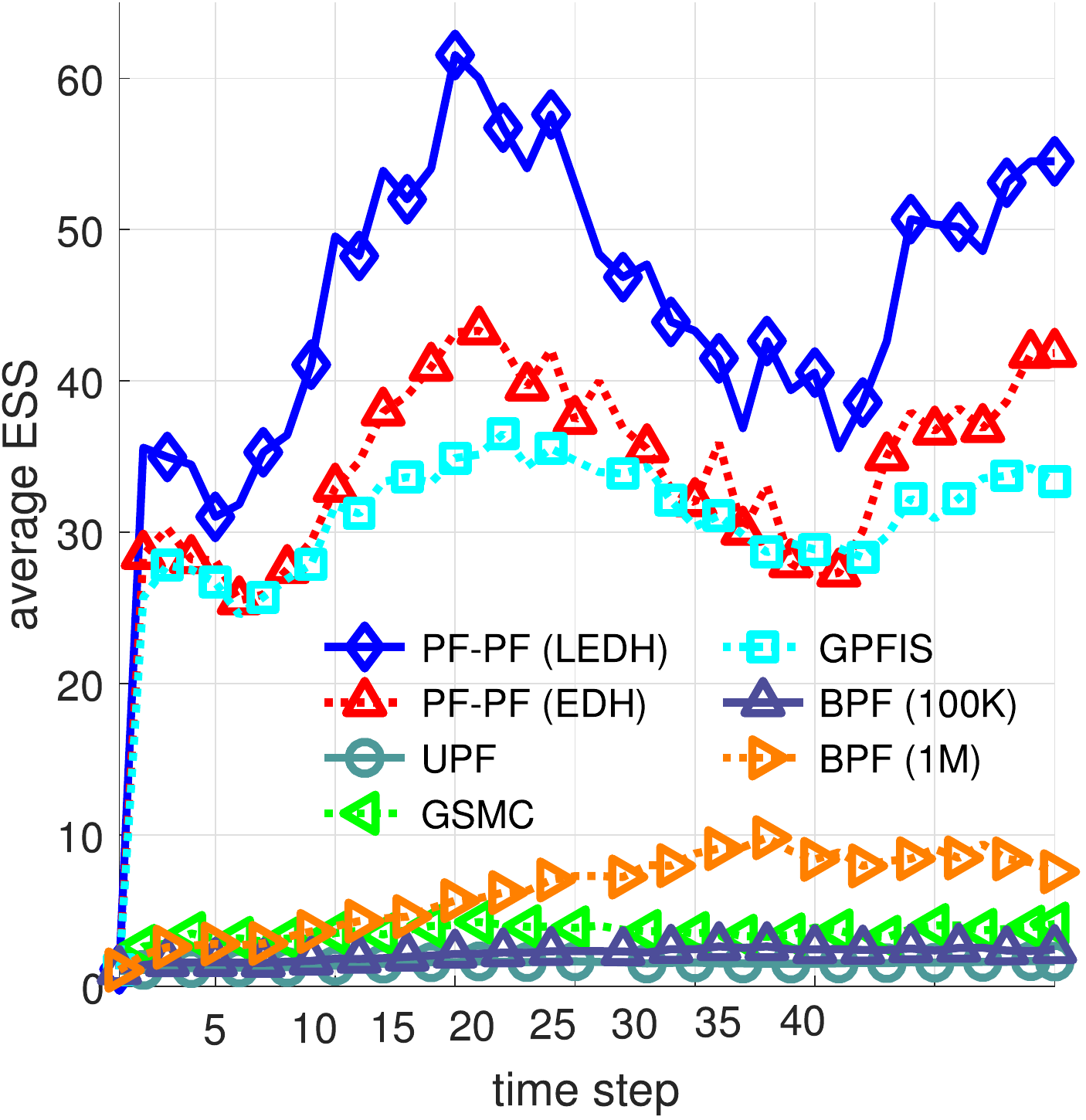}
 \caption{\label{fig:ESS_acoustic_average}
 Average effective sample size at each time step in the multi-target acoustic tracking example.}
\end{figure}

From Table~\ref{tab:time_error_acoustic}
and Figure~\ref{fig:ESS_acoustic_average},
we can see that the PF-PF algorithms and GPFIS
provide the highest effective
sample sizes among all tested algorithms with importance sampling.
Even with one million particles, the average ESS of the bootstrap particle filter is still less than 7, and the cost of computation is
more than $3$ times the cost of the PF-PF (LEDH).
The UPF has relatively high computational cost, as unscented
transformations are performed for each sigma point,
and the number of sigma points for each particle is proportional
to the state dimension.
The GSMC algorithm has a small effective sample size of 3.5. This shows that
the transport map constructed by the ESRF is not effective
for GSMC in this example.
The PF-PF (LEDH) has a small increase of execution time compared with the LEDH,
showing that the weight update is very efficient.

\subsection{Large spatial sensor networks: linear Gaussian example}
\label{sec:simulation_linear_Gaussian}

\subsubsection{Simulation setup}
\label{sec:setup_linear_Gaussian}

To compare the accuracy of proposed filters relative to
theoretical optimal accuracy, we examine the filters'
performance in a simple linear Gaussian filtering problem with
the spatial sensor network setup proposed in~\cite{septier2016}.
There are $d$ sensors deployed uniformly on a two-dimensional
grid $\{1,2,\ldots,\sqrt{d}\}\times\{1,2,\ldots,\sqrt{d}\}$, and
$d$ is set to $64$ in this example.
Each sensor collects measurements, independently of the other sensors,
about the underlying state at its physical location. Denote the state
at the $c$-th sensor's position at time $k$ by $x_k^c \in \mathbb{R}$,
and its measurement as $z_k^c \in \mathbb{R}$. Then the full state
at all sensor positions at time $k$ is denoted by
$x_k = [x_k^1, x_k^2, \ldots, x_k^d]' \in \mathbb{R}^d$,
and all measurements at time $k$ form the measurement vector
$z_k = [z_k^1, z_k^2, \ldots, z_k^d]' \in \mathbb{R}^d$.

The dynamic model and the measurement model are, respectively:
\begin{align}
x_k &= \alpha x_{k-1} + v_k\,,\\
z_k &= x_k + w_k\,,
\end{align}
where $\alpha = 0.9$, and $w_k \in \mathbb{R}^d$ is a zero-mean
Gaussian random vector with covariance
$\Sigma_z = \sigma_z^2 \mathbb{I}_{d\times d}$.
$v_k \in \mathbb{R}^d$ is a zero-mean
Gaussian random vector with covariance $\Sigma$.
The $(i,j)$-th entry of $\Sigma$ is 
\begin{align}
\Sigma_{i,j} = \alpha_0 e^{-\frac{||R^i-R^j||_2^2}{\beta}}+\alpha_1\delta_{i,j}\,,
\label{eq:dispersion}
\end{align}
where $||\cdot||_2$ is the Euclidean norm,
$R^i \in \mathbb{R}^2$ is the physical position
of sensor $i$, and $\delta_{i,j}$
is the Kronecker delta symbol ($\delta_{i,i}=1$ and $\delta_{i,j}=0$ for $i\neq j$).  This equation implies that the
noise dependence increases when the spatial distance between two sensors decreases.

Following~\cite{septier2016}, we set $\alpha_0 = 3, \alpha_1 = 0.01,
\beta = 20$.  We vary the value of $\sigma_z$, to investigate
the sensitivity of algorithms with respect to the level of measurement noise.
All true states start with $x_0^c = 0$, for $c = 1, \ldots, d$.
The experiment is executed 100 times for 10 time steps.

\subsubsection{Parameter values for the filtering algorithms}
\label{sec:param_value_linear_Gaussian}

We compare the proposed particle flow particle filter (PF-PF) algorithms
with various Kalman-type filters and other particle filters
with different measurement noise levels.
In the linear Gaussian scenario, the Kalman filter (KF)
provides the exact posterior distribution, thus
it gives the optimal filtering accuracy.
The EDH, the ESRF, and the UKF are all derived based
on the linear Gaussian assumptions. They thus provide
near-optimal filtering performance, although their
use of Monte Carlo samples or sigma points make
their solutions deviate slightly from optimal.
The BPF is the vanilla particle filter that uses
the dynamic model to propose particles. The main goal
of this experiment is to compare the PF-PF algorithms
with those filters that provide optimal or near-optimal performance.
We also demonstrate the challenges to particle filters
from highly informative measurements by varying the measurement
noise level.

For the algorithms employing particle flow, the step sizes are set to
be the same as those reported in Section~\ref{sec:param_acoustic}
and the Kalman filter covariance equations are used to estimate the predicted covariance. All filters are initialized with the same true
state $0$ in each state dimension.

\subsubsection{Experimental results}

Table~\ref{tab:linear_Gaussian}
reports the average mean squared errors (MSEs) over
100 simulation trials and the execution
times per time step with $\sigma_z$ set to $2$, $1$, or $0.5$.

\begin{table}[htbp]
\small
\centering
\tabcolsep=0.043cm
\caption{Average MSE, ESS and execution time per step
in a 64-dimensional linear Gaussian example of the large spatial sensor networks simulation with different measurement noise levels, among
100 simulation trials. Results are produced with an Intel i7-4770K 3.50GHz CPU and 32GB RAM.}
\label{tab:linear_Gaussian}
\renewcommand{\arraystretch}{2}
\scalebox{0.9}
{
\begin{tabular}{|c|c||c|c||c|c||c|c||c|}
	\hline
    \multicolumn{2}{|c||}{$\sigma_z$} &  \multicolumn{2}{c||}{2} &  \multicolumn{2}{c||}{1} & \multicolumn{2}{c||}{0.5} & \multirow{2}{*}{\specialcell{Exec.\\time (s)}}\\\cline{1-8}
Algorithm & \specialcell{Particle\\num.} & 
\specialcell{Avg.\\MSE} & \specialcell{Avg.\\ESS} & 
\specialcell{Avg.\\MSE} & \specialcell{Avg.\\ESS} & 
\specialcell{Avg.\\MSE} & \specialcell{Avg.\\ESS} &
\\\hline
PF-PF (LEDH) & 200 & 0.61 & 28 & 0.25 & 23 & 0.10 & 19 & 1.8\\\hline
PF-PF (EDH) & 200 & 0.62 & 28 & 0.26 & 23 & 0.11 & 19 & 0.01\\\hline
PF-PF (EDH) & $10^4$ & 0.53 & 1118 & 0.22 & 973 & 0.09 & 830 & 0.24\\\hline
EDH & 200 & 0.49 & N/A & 0.19 & N/A & 0.07 & N/A & 0.01\\\hline
KF & N/A & 0.49 &  N/A & 0.18 & N/A & 0.07 & N/A & 0.0005\\\hline
UKF &  N/A & 0.49 & N/A & 0.18 & N/A & 0.07 & N/A & 0.007\\\hline
UPF & 200 & 0.87 & 22 & 0.32 & 22 & 0.12 & 22 & 1.4 \\\hline
ESRF & 200 & 0.50 & N/A & 0.19 & N/A & 0.07 & N/A & 0.001\\\hline
GSMC & $200$ & 0.52 & 2.0 & 0.19 & 2.0 & 0.08 & 2.0 & 0.002 \\\hline
BPF & $200$ & 1.20 & 1.9 & 1.1 & 1.2 & 1.1 & 1.0 & 0.00006 \\\hline
BPF & $10^5$ & 0.54 & 49 & 0.32 &3.2 & 0.29 & 1.3 & 0.24\\\hline
\end{tabular}}
\end{table}

We observe that the Kalman filter, the ESRF, the UKF, and the EDH filter have the smallest mean squared errors for different $\sigma_z$ values.
This is expected as the linear Gaussian models
match their model assumptions.
We also observe that with 200 particles,
the GSMC algorithm has the smallest average MSEs
among all particle filters.
When the number of particles is increased to $10^4$,
the PF-PF (EDH) is still relatively efficient,
and has average MSEs close to those of the Kalman filter.

We also note that for the BPF, as $\sigma_z$
gets smaller, the average ESS drastically decreases,
which indicates that smaller measurement noise leads to
a more challenging situation for many particle filters.
However, for the PF-PF algorithms, the average effective
sample size only decreases slightly as $\sigma_z$ decreases
from 2 to 0.5,
showing that particle flows are able to propagate particles into the region
of high posterior densities, even in the more challenging
scenarios with highly informative measurements.
When $\sigma_z = 0.5$, the PF-PF (EDH) and the GSMC with 200 particles
are more accurate than the BPF with $10^5$ particles.

We next investigate the sensitivity of the proposed algorithm to the approximation
of the predictive covariance. We add noise to the
Kalman filter-estimated
covariance $\hat{P}$ to generate a noisy estimate $\tilde{P}$.
To ensure that the noise-injected $\tilde{P}$ is positive definite 
(as a requirement for the covariance matrix), we first perform 
eigendecomposition of $\hat{P}$ and obtain the eigenvalues
$D \in \mathbb{R}^d$ and the corresponding right
eigenvectors $V \in \mathbb{R}^{d\times d}$.
We then generate $\tilde{D} = \xi \circ D$,
where $\xi \in \mathbb{R}^d$ is a random vector
and $\circ$ denotes the dot product. The elements of $\xi$
are independent and are distributed according to a log-normal distribution
$\mbox{ln} N(0, \sigma_p^2)$. Because $\hat{P}$ is positive definite,
each element of $D$ is positive. All elements
of $\epsilon$ are also positive as they are generated from a log-normal
distribution. Thus, all elements of $\tilde{D}$ are positive.
We then generate $\tilde{P} = V \times \mbox{diag}(\tilde{D}) \times V^T$
where $\mbox{diag}(D)$ denotes a diagonal matrix
with the elements of the vector $D$ on the diagonal.
With this procedure, $\tilde{P}$ is positive definite as all
of its eigenvalues are positive. Numerical results show that the expected value of $\frac{|\hat{P}-\tilde{P}|}{|\hat{P}|}$, the ratio between the Euclidean norm of the error added to $\hat{P}$
and the Euclidean norm of $\hat{P}$, is approximately equal to $\sigma_p$. We report in Table~\ref{tab:sensitivity} 
the MSE from 100 simulation trials with different values of $\sigma_p$.

\begin{table}[htbp]
\small
\centering
\tabcolsep=0.03cm
\caption{Average MSE and ESS in the 64-dimensional linear Gaussian example of the large 
spatial sensor networks with $\sigma_z = 1$ and different predictive
covariance estimation errors, among 100 simulation trials.}
\label{tab:sensitivity}
\renewcommand{\arraystretch}{2}
\scalebox{0.82}
{
\begin{tabular}{|c|c|c|c|c|c|c|}
	\hline
	 Algorithm & \multicolumn{2}{c|}{PF-PF (LEDH)} & \multicolumn{2}{c|}{PF-PF (EDH)}  & \multicolumn{2}{c|}{PF-PF (EDH)} \\[1ex]\hline
     Particle num. & \multicolumn{2}{c|}{200}& \multicolumn{2}{c|}{200} & \multicolumn{2}{c|}{$10^4$} \\ \hline
    & Avg. MSE & Avg. ESS & Avg. MSE & Avg. ESS & Avg. MSE & Avg. ESS \\\hline
   $\sigma_p = 0$ & 0.25 & 23 & 0.26 & 23 & 0.22 & 973 \\\hline
$\sigma_p = 0.1$ & 0.39 & 26 & 0.39 & 26 & 0.34 & 1018 \\\hline
$\sigma_p = 0.2$ & 0.40 & 23 & 0.40 & 23 & 0.35 & 894 \\\hline
$\sigma_p = 0.5$ & 0.43 & 12 & 0.43 & 12 & 0.39 & 343 \\\hline
$\sigma_p = 1$ & 0.54 & 3.6 & 0.54 & 3.6 & 0.51 & 32 \\\hline
\end{tabular}}
\end{table}

We observe that as we increase the amount of noise added into
the estimated covariance, by increasing the standard deviation
$\sigma$ of the log-normal distribution, the PF-PF (EDH) tends
to have higher mean squared error (MSE) and smaller effective sample size (ESS).
However, even when $\sigma_p = 0.2$, which adds considerable noise to the eigenvalues of $P$, there is only a minor reduction in the ESS. It is only when
$\sigma_p$ reaches 0.5 or 1, indicating a very noisy predictive covariance
matrix estimate, that the effective sample size decreases significantly. 

Khan et al. investigates the effects of different covariance approximation techniques on the  performance of the particle flow filter in~\cite{khan2015}. Our experience with these techniques is that their impact on the overall tracking performance depends on the specific nature of the application. For consistency across our simulation results, we use the EKF covariance equations for the estimation of predictive covariance.

\subsection{Large spatial sensor networks: Skewed-t dynamic model
and count measurements}

\subsubsection{Simulation setup}
\label{sec:setup_large_sensor_networks}


The spatial placement of $d$ sensors is the same as 
that introduced in~\ref{sec:setup_linear_Gaussian}.
The dynamic model of the underlying state $x_k$ in this
example follows the multivariate Generalized Hyperbolic (GH)
skewed-t distribution, which is a heavy-tailed distribution
that is useful for modelling physical processes and financial markets
with extreme behavior and asymmetric data~\cite{zhu2010}. We have
\begin{align}
p(x_k|&x_{k-1}) = K_{\frac{\nu+d}{2}}(\sqrt{(\nu+Q(x_k))(\gamma^T\Sigma^{-1}\gamma)})\nonumber\\
\hspace{-0.25cm}\times &\dfrac{e^{(x_k-\mu_k)^T\Sigma^{-1}\gamma}}{\sqrt{(\nu+Q(x_k))(\gamma^T\Sigma^{-1}\gamma)}^{-\frac{\nu+d}{2}}(1+\frac{Q(x_k)}{\nu})^{\frac{\nu+d}{2}}} \,,
\label{eq:mGH}
\end{align}
where $K_{\frac{\nu+d}{2}}$ is the modified Bessel function 
of the second kind of order $\frac{\nu+d}{2}$,
$\mu_k = \alpha x_{k-1}$,
$Q(x_k) = (x_k-\mu_k)^T\Sigma^{-1}(x_k-\mu_k)$,
and the $(i,j)$-th entry of $\Sigma$ is again defined by Equation~\eqref{eq:dispersion}.
The parameters $\gamma$ and $\nu$
determine the shape of the distribution.
The covariance is given by:
\begin{align}
\tilde{\Sigma} = \dfrac{\nu}{\nu-2}\Sigma
+\dfrac{\nu^2}{(2\nu-8)(\frac{\nu}{2}-1)^2}\gamma \gamma^T\,.
\end{align}

The measurements are count data with the following Poisson distribution
\begin{align}
p(z_k|x_k) = \prod_{c=1}^{d}\mathcal{P}_0(z_k^c;m_1 e^{m_2 x_k^c}) \,,
\end{align}
where $\mathcal{P}_0(\cdot;m)$ is the Poisson$(m)$ distribution.
We set $m_1 = 1$ and $m_2 = \frac{1}{3}$. $d$ is set to $144$ or $400$
to represent two high-dimensional filtering scenarios.
Again, each scenario is executed 100 times for 10 time steps.
We choose the parameter values of the simulation setup
to be the same as those used in~\cite{septier2016},
because we would like to evaluate
the proposed filters in the same simulation setups where
the state-of-the-art SmHMC algorithm has been compared to other Langevin and Hamiltonian-based algorithms and exhibited the
smallest estimation errors.

\subsubsection{Parameter values for the filtering algorithms}

We compare the proposed PF-PF algorithms
with the Kalman-type filters and particle filters evaluated
in Section~\ref{sec:simulation_linear_Gaussian}. In addition,
we evaluate two filters specifically designed for high-dimensional
nonlinear filtering: one is the SmHMC, which exhibits the smallest mean squared error
(MSE) in~\cite{septier2016}, and the other is the block particle filter~\cite{rebeschini2015}.
We do not compare with the GPFIS
algorithm~\cite{bunch2016}, due to its prohibitively large
computational cost.

Parameter values for the SmHMC algorithm
are set to those identified in~\cite{septier2016}, 
including the number of particles which is 200. We evaluate
the performance for the PF-PFs using 200 particles, as well as a higher number of particles for the PF-PF (EDH) with the constraint that its computational time remains
less than that of SmHMC with 200 particles. 
The step sizes of the particle flow-type algorithms are set to
be the same as those reported in Section~\ref{sec:param_acoustic}.
Since the measurement noise depends on the state,
the measurement covariance $R$ is updated in each
discretized particle flow step and before the EKF update. For the PF-PF (LEDH),
$R$ is updated using $\bar{\eta}^i$ for
each particle; for the PF-PF (EDH), $\bar{\eta}$
is used. All filters are initialized with the same true
state $0$ in each state dimension, which is the scenario
explored in~\cite{septier2016}.

\subsubsection{Experimental results}

Table~\ref{tab:time_error_large_sensor_networks_delta_prior}
reports the average mean squared errors (MSEs) over
100 simulation trials and the execution
times per time step. We observe that the EDH and the LEDH filters
have very similar average MSE errors. This suggests that
computing the flow parameters
separately for each particle does not provide additional gains in this
setting, which is different from the result shown
in Section~\ref{sec:acoustic}. Thus, the EDH is preferred over the LEDH as it is much less computationally demanding.

\begin{table}[htbp]
\small
\centering
\tabcolsep=0.043cm
\caption{Average MSE, ESS and execution time per step
in the large spatial sensor networks simulation
with a Skewed-t dynamic model and count measurements.
Results are produced with an Intel i7-4770K 3.50GHz CPU and 32GB RAM. 
When there is a parenthesis after the average MSE values,
it indicates the number
of lost tracks out of 100 simulation trials where
the lost tracks are defined as those whose
average estimation errors are greater than $\sqrt{d}$.
The average MSE is calculated
with the simulation trials where tracking is not lost.}
\label{tab:time_error_large_sensor_networks_delta_prior}
\renewcommand{\arraystretch}{2}
\scalebox{0.9}
{
\begin{tabular}{|c|c||c|c|c||c|c|c|}
	\hline
    \multicolumn{2}{|c||}{d} &  \multicolumn{3}{c||}{144} &  \multicolumn{3}{c|}{400}\\\hline
Algorithm & \specialcell{Particle\\num.} & 
\specialcell{Avg.\\MSE} & \specialcell{Avg.\\ESS} & 
\specialcell{Exec.\\time (s)} & \specialcell{Avg.\\MSE} &
\specialcell{Avg.\\ESS} & \specialcell{Exec.\\time (s)}\\\hline
PF-PF (LEDH) & 200 & 0.95 & 6.7 & 7.8 & 1.04 & 3.4 & 110 \\\hline
PF-PF (EDH) & 200 & 0.96 & 6.6 & 0.05 & 1.05 & 3.4 & 0.5 \\\hline
PF-PF (EDH) & $10^4$ & 0.82 & 81 & 1.6 & 0.89 & 20 & 4.8 \\\hline
LEDH & 200 & 0.71 & N/A & 6.8 & 0.62 & N/A & 88 \\\hline
EDH & 200 & 0.69 & N/A & 0.05 & 0.60 & N/A & 0.5 \\\hline
EDH & $10^4$ & 0.69 & N/A & 0.6 & 0.60 & N/A & 2.5 \\\hline
SmHMC & 200 & 0.83 & N/A & 15 & 0.73 & N/A & 87 \\\hline
Block PF & $10^4$ & 1.7 & N/A & 10 & 1.6 & N/A & 29 \\\hline
EKF & N/A & 2.5 (29) &  N/A & 0.002 & 3.4 (18) & N/A & 0.03\\\hline
UKF &  N/A & 2.4 (34) & N/A & 0.05 & 3.8 (27) & N/A & 1.2\\\hline
UPF & 200 & 2.2 (34) & 3.0 & 12 & 4.6 (43) & 1.4 & 236 \\\hline
ESRF & 200 & 2.3 (23) & N/A & 0.01 & 2.8 (15) & N/A & 0.05\\\hline
GSMC & $200$ & 2.4 (22) & 1.1 & 0.02 & 3.5 (23) & 1.0 & 0.06\\\hline
BPF & $10^5$ & 1.8 & 6.3 & 0.7 & 4.0 (1) & 1.3 & 2.3 \\\hline
BPF & $10^6$ & 1.3 (1) & 26 & 6.8 &3.3 (1) & 1.5 & 23\\\hline
\end{tabular}}
\end{table}

We also observe that the PF-PF (EDH) and the
PF-PF (LEDH) lead to larger average
MSEs than the EDH or the LEDH
with the same number of particles.
The EDH and LEDH filters use particle flow to generate approximations of the posterior distribution; the PF-PF algorithms perform subsequent importance sampling to modify this approximation. The importance sampling makes the filter statistically consistent, but in high dimensions, it can introduce a high variance in the weights, leading to poorer performance in state estimation.
This sampling error can be reduced
by increasing the number of particles, and
we see that with 10000
particles, the average MSE of PF-PF (EDH) has
significantly decreased, as the effective sample
size increases considerably.
Even with 10000 particles, the PF-PF (EDH) is more
computationally efficient than the SmHMC with 200 particles. The estimation errors
are similar for the PF-PF (EDH) and SmHMC
when $d=144$, although the average error of the PF-PF (EDH) is considerably higher when $d=400$.
The Block PF has relatively large estimation errors,
possibly due to the intrinsic bias introduced from
the blocking step as stated in~\cite{rebeschini2015}.
The Kalman-type filters frequently struggle to track the state at all,
leading to lost tracks. The ESRF in particular has a high number
of lost tracks when the state dimension is 400, as the sample predictive
covariance is often close to singular. The GSMC method and the UPF,
which use the ESRF and the UKF, respectively to construct the proposal distributions, also exhibit poor performance. Even with a million particles, the BPF performs relatively poorly compared to most other algorithms.

We evaluate the effective sample size (ESS) for the PF-PF algorithms
and other particle filters. The standard ESS estimate is not meaningful for
SmHMC, because due to its MCMC structure there are no weights associated with
the particles; for the block PF we can only calculate an ESS for each block,
so the value is not comparable. The ESS values indicate why the PF-PF filters
perform worse in this setting compared to the acoustic tracking example.
Only a very small fraction of the particles have a significant weight.
There is however, a substantial improvement compared to the BPF, the UPF
and the GSMC;
the BPF requires 100 times more particles to achieve comparable ESS values.

\section{Conclusions}
\label{sec:conclusion}

In this paper, we have presented particle flow particle filtering algorithms
with efficient importance weight computation.
We proved that the embedded particle
flows possess the invertible mapping property, which is crucial for achieving
straightforward weight updates. The weight updates add only a small computation cost
to particle flow filters they build upon.

We have evaluated the proposed algorithms' performance in
three scenarios. In the multi-target tracking simulation setup,
the PF-PF (LEDH) with 500 particles leads to the smallest tracking error 
and maintains particle clouds with the highest effective sample size at 
most time steps. This demonstrates that the PF-PF (LEDH) is capable of producing
better particle representations of posterior distributions
than other filtering algorithms with much higher 
computational cost in this highly informative measurement setting.
In a linear Gaussian filtering example, we show that the PF-PF algorithms
approach the optimal accuracy with a reasonable number of
particles in different settings with various levels of measurement noise.
We also discuss the sensitivity of the PF-PF with respect
to the estimation errors of the predictive covariance.
In the large spatial sensor network setting
where the state dimension is high and the models are non-Gaussian,
the EDH filter provides the smallest average MSE
and is computationally efficient.
The error introduced by incorporating importance sampling
in the proposed PF-PF (EDH) algorithm outweighs the approximation error in the 
EDH filter.

The proposed PF-PF algorithms are computationally efficient particle filters
that can perform well in high-dimensional settings, but the last simulation
motivates the development of improved mechanisms for using the particle flow 
procedures to construct a consistent filter.
An important future research direction is the construction of improved invertible
particle flows, which may be achieved by employing stochastic particle 
flows~\cite{deMelo2015,daum2013a} or performing linearization at the actual particle locations. It can be significantly more difficult to construct
stochastic particle flows with the invertible mapping property,
but the diffusion of particles may lead to more diverse
sets of particles and hence better proposal distributions.
Other directions include the identification of
new particle flows based on mixture models
to allow the construction of proposal distributions that can match
multi-modal distributions, the incorporation of iterative importance sampling ~\cite{morzfeld2016} to increase the effective sample size
in high-dimensional filtering scenarios,
the integration of quasi Monte
Carlo~\cite{gerber2015} with particle flow particle filters
for faster convergence,
and convergence studies of particle filters with invertible
particle flow.

\section*{Acknowledgment}

The authors would also like to thank Fran\c{c}ois Septier and
Gareth W. Peters for making Matlab Codes associated to~\cite{septier2016}
publicly available. This work was conducted with the support of the Natural Sciences and Engineering Research Council of Canada (NSERC 260250).

\ifCLASSOPTIONcaptionsoff
  \newpage
\fi

\bibliographystyle{IEEEtran}
\bibliography{PFPF_jrnl_2017}

%
\begin{IEEEbiography}[{\includegraphics[width=1in,height=1.25in,clip,keepaspectratio]{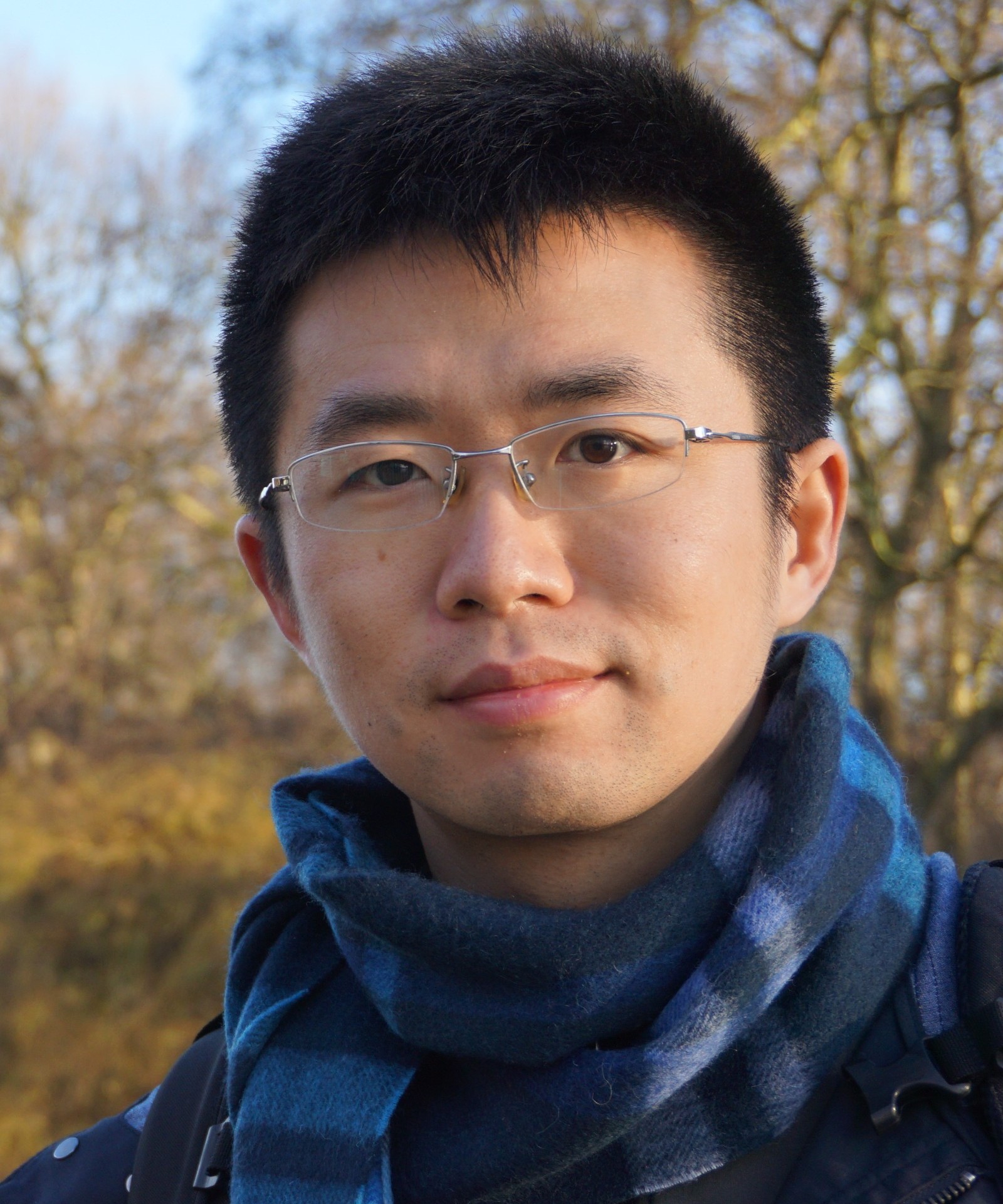}}]{Yunpeng Li} (S'16)
is a Ph.D. candidate at the Department of Electrical and Computer Engineering at McGill University, Canada. He received the B.A. and M.S. Eng. degrees from the Beijing University of Posts and Telecommunications, China, in 2009 and 2012, respectively.
He joined the Machine Learning Research Group at University of Oxford, U.K., as a Postdoctoral Researcher in Machine Learning in April 2017.
He has conducted research internships at HP Labs China (Spring 2012) and McGill University (Summer 2011, Summer 2010). His research interests include Monte Carlo methods in high-dimensional spaces, microwave breast cancer detection, and Bayesian inference.
\end{IEEEbiography}

\begin{IEEEbiography}[{\includegraphics[width=1in,height=1.25in,clip,keepaspectratio]{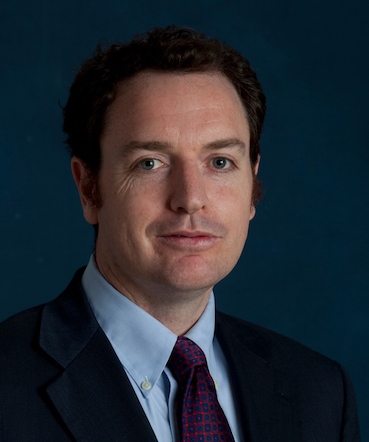}}]{Mark Coates} (SM'04)
received the B.E. degree in computer systems engineering from the University of Adelaide, Australia, in 1995, and a Ph.D. degree in information engineering from the University of Cambridge, U.K., in 1999. He joined McGill University (Montreal, Canada) in 2002, where he is currently an Associate Professor in the Department of Electrical and Computer Engineering. He was a research associate and lecturer at Rice University, Texas, from 1999-2001. In 2012-2013, he worked as a Senior Scientist at Winton Capital Management, Oxford, UK. He was an Associate Editor of IEEE Transactions on Signal Processing from 2007-2011 and a Senior Area Editor for IEEE Signal Processing Letters from 2012-2015. In 2006, his research team received the NSERC Synergy Award in recognition of their successful collaboration with Canadian industry, which has resulted in the licensing of software for anomaly detection and Video-on-Demand network optimization. Coates' research interests include communication and sensor networks, statistical signal processing, and Bayesian and Monte Carlo inference. 
\end{IEEEbiography}

\end{document}